\documentclass[a4paper,aps,prd,10pt,showpacs,twocolumn,nobibnotes,nofootinbib]{revtex4-1}
\usepackage{graphicx,amsmath,amsfonts,amssymb,revsymb,dcolumn,epsfig,bm,xspace}
\usepackage{hyperref}
\usepackage[utf8]{inputenc}
\usepackage{color}
\usepackage{bm}
\usepackage{mathtools}

\newcommand{\OO}[1]{\mathcal{O}\left(#1\right)}
\newcommand{\dd}{\mathrm{d}}
\newcommand{\lie}{\pounds}
\newcommand{\del}{\partial}
\newcommand{\dirac}[2]{\delta^{#1}(#2)}
\newcommand{\Poisson}[2]{\left\{#1,\;#2\right\}}
\newcommand{\Comm}[2]{\left[#1,\;#2\right]}

\newcommand{\Prod}[2]{\left(#1, #2\right)}
\newcommand{\ci}{\mathsf{i}}
\newcommand{\Prp}{\mathsf{P}}
\newcommand{\CS}{\mathsf{J}}
\renewcommand{\Im}{\operatorname{Im}}
\renewcommand{\Re}{\operatorname{Re}}
\newcommand{\myfrac}[2]{%
    \setbox0\hbox{$#1$}        
    \dimen0=\wd0               
    \setbox1\hbox{$#2$}        
    \dimen1=\wd1               
    \ifdim\wd0<\wd1            
        \dfrac{#1\hfill}{#2}   
    \else                      
        \dfrac{#1}{#2\hfill}   
    \fi
}
\newcommand{\g}{g}
\newcommand{\n}{n}
\newcommand{\h}{\gamma}
\newcommand{\fint}{\sqrt{-g}\dd^4x}

\newcommand{\tint}{\mathbf{e}^3}
\newcommand{\tintz}{\tilde{\mathbf{e}}^3}
\newcommand{\ddt}{\dd{}t}
\newcommand{\ddx}{\dd^3x}
\newcommand{\fintc}{\varepsilon}
\newcommand{\cd}{\nabla}
\newcommand{\ST}{\Sigma}
\newcommand{\EC}{\mathcal{K}}
\newcommand{\EX}{\Theta}
\newcommand{\mf}{\phi}
\newcommand{\pmf}{\Pi_\mf}
\newcommand{\dpmf}{\dot{\Pi}_\mf}
\newcommand{\mvo}{\hat{\chi}}
\newcommand{\fsmvo}{\tilde{\chi}}
\newcommand{\mfo}{\hat{\mf}}

\newcommand{\fsmfo}{\tilde{\mf}}
\newcommand{\fspmfo}{\tilde{\Pi}_\mf}
\newcommand{\fsmf}{\bar{\mf}}

\newcommand{\wf}[2]{\Psi_{#1}\left[#2\right]}
\newcommand{\gH}{\mathcal{H}}

\newcommand{\scd}{D}
\newcommand{\HF}[1]{\mathcal{Y}_{#1}}
\newcommand{\EV}[1]{\vert\HFi\vert}

\newcommand{\HFi}{\bm{k}}
\newcommand{\BR}{\mathsf{R}}
\newcommand{\BE}{\mathsf{E}}
\newcommand{\BRi}{\mathsf{r}}

\newcommand{\BSi}{\mathsf{s}}
\newcommand{\BT}{\mathsf{T}}
\newcommand{\BTi}{\mathsf{t}}

\newcommand{\BF}[1]{\mathsf{U}_{#1}}
\newcommand{\dBF}[1]{\dot{\mathsf{U}}_{#1}}
\newcommand{\dBT}{\dot{\mathsf{T}}}
\newcommand{\SEV}{e}
\newcommand{\AO}{\mathsf{a}}

\newcommand{\AOh}{\mathsf{A}}
\newcommand{\bra}[1]{\left\langle#1\right\vert}
\newcommand{\ket}[1]{\left\vert{#1}\right\rangle}
\newcommand{\braket}[2]{\left\langle#1\middle\vert#2\right\rangle}
\newcommand{\braketOP}[3]{\left\langle#1\middle\vert#2\middle\vert#3\right\rangle}
\newcommand{\ofreq}{\nu}
\newcommand{\om}{m}
\newcommand{\omfreq}{\xi}
\newcommand{\onm}[1]{\om_{c_{#1}}}
\newcommand{\onfreq}[1]{\ofreq_{c_{#1}}}
\newcommand{\onmfreq}[1]{\omfreq_{c_{#1}}}
\newcommand{\donmfreq}[1]{\dot{\omfreq}_{c_{#1}}}

\newcommand{\SM}{\mathbb{S}}

\newcommand{\lapse}{N}

\DeclareFontFamily{U}{mathx}{\hyphenchar\font45}
\DeclareFontShape{U}{mathx}{m}{n}{<-> mathx10}{}
\DeclareSymbolFont{mathx}{U}{mathx}{m}{n}
\DeclareMathAccent{\widebar}{0}{mathx}{"73}

\begin{document}
\title{Unitary evolution, canonical variables and vacuum choice for general quadratic Hamiltonians in spatially homogeneous and isotropic space-times}
\author{Sandro~D.~P.~Vitenti}
\email{dias@iap.fr}
\affiliation{CAPES Foundation, Ministry of Education of Brazil, Bras\'{\i}lia -- DF 70040-020, Brazil}
\affiliation{${\mathcal G}\mathbb{R}\varepsilon\mathbb{C}{\mathcal O}$ -- Institut d'Astrophysique de Paris, UMR7095 CNRS, Universit\'e Pierre \& Marie Curie, 98 bis boulevard Arago, 75014 Paris, France}

\date{\today}

\begin{abstract}
Quantization of arbitrary free scalar fields in spatially homogeneous and isotropic space-times is considered. The quantum representation allowing a unitary evolution for the fields is taken as a requirement for the theory. Studying the group of linear canonical transformations, we show the relations between unitary evolution and choice of canonical variables. From these relations we obtain the conditions on the Hamiltonian such that there are canonical variables for which the field has unitary evolution. We then compute the linear transformation leading to these variables, also proving that they are unique. We obtain these results by developing the asymptotic analysis of the fields using the action angle variables, which proves to be a generalization of the usual Wentzel-Kramers-Brillouin approximation. These tools allow us to re-frame the adiabatic vacuum condition in a extensible format by using the action angle variables to relate these vacuum choices to those where the particle number density does not depend on angle (fast) variables. Finally, we develop a larger set of canonical variables relating the adiabatic vacuum conditions with the smearing of the quantum fields. This set of canonical variables also connects the adiabatic vacuum conditions with the instantaneous Hamiltonian diagonalization vacuum choice. 
\end{abstract}


\maketitle

\section{Introduction}

Differently from the Quantum Mechanics (QM), Quantum Field Theory (QFT) has an additional structure necessary for its development, the representation choice for the field. Given the fact that the QFT has an infinity number of degrees of freedom, the Stone-Von-Neumann theorem~\cite{Reed1979} no longer guarantees that any choice of representation will be unitarily equivalent to any other. This is physically irrelevant in QM and, thus, one can choose the most computational suitable representation for each problem.  Consequently, this additional ambiguity presented in QFT must be solved by imposing some physical requirement. In particular, on the Minkowsky space-time, QFT has a natural choice of representation induced by the Poincar\'{e} symmetry group. The existence of a symmetry group also constrains in a non-explicit way the choice of canonical variables used to represent the field. 

Given a free scalar field on Minkowsky space-time, for example, with a canonical kinetic term and time independent mass, its Hamiltonian will be time-independent. Any time-dependent re-scaling of the field would result in a new time-dependent Hamiltonian, which is neither conserved nor coincides with the energy of the system. In other words, there is a special field variable where the Hamiltonian is time independent. For more general backgrounds, when there is no symmetry group aiding the choice of representation for the field, one also does not have a clear guideline to define the appropriated canonical variable to represent the field. Particularly, when there is no time translation symmetry, the Hamiltonian will potentially be time-dependent. In this case, any re-scaling of the field would just lead to another time-dependent Hamiltonian.

The problem of the choice of canonical variables is more evident in the treatment of the cosmological perturbations. In this setting the first order perturbations will be described by a constrained Hamiltonian system, quadratic on the fields. Although, differently from the scalar field, the canonical variables in the non-constrained Hamiltonian are the result of the solution of the original constrained Hamiltonian system. In other words, different solutions of the constraints lead to different canonical variables in the non-constrained Hamiltonian. For example, in~\cite{Vitenti2013} we are lead to the curvature perturbation $\zeta$ as the canonical variable representing the perturbations when we solve the constraints. However, this is an arbitrary choice and one could choose any other variable re-scaling $\zeta$ by a simple background quantity. 

In principle, one could expect that the ambiguity in choosing a canonical variable to represent the field would be of no interest. However, in a series of works~\cite{Torre2002,Corichi2006,Cortez2007,Corichi2007,BarberoG2008,Vergel2008,Cortez2011,Gomar2012,Cortez2012,Fernandez-Mendez2012,Cortez2013} it was shown that there is an important interplay between the choice of canonical variables and the unitary implementability of the field operators time evolution. These works extend the pioneering results of \citet{Parker1968,Parker1969,Parker2012}, in which he showed that the time evolution is not unitarily implemented and, introduced the adiabatic regularization to remove the divergent part of the number operator. However, those mentioned works show that the time evolution is indeed unitary for a special choice of field variable. This apparent contradiction comes from the fact that, in the latter cited works, the author only considered one canonical variable to represent the field. In the references above, the authors show for many different cases that, re-scaling the field, one can quantize the modified version of the field obtaining an unitary evolution.

In a recent work~\cite{Agullo2015} the authors advocate for a new definition of unitary evolution. This concept generalization matches the prescription discussed above, where one re-scales the field to obtain the unitary evolution in the standard sense. This new definition of unitary evolution can, therefore, be used to justify the QFT in the original variables. In this work we will always refer to the unitary evolution in the standard sense.

The purpose of this paper is fourfold. First, we exhibit the requirement on the Hamiltonian of free fields for which one can find a set of canonical variables and its representation, where the time evolution is unitarily implemented. For this, we use a time-dependent Linear Canonical Transformation (LCT) and consider background geometries with homogeneous spatial hypersurfaces. Following, as our second result, we show explicitly which are these canonical variables. The results apply to systems in which the Lagrangian (and, consequently, the Hamiltonian) is quadratic on the fields. There are two main categories described by such systems: free fields evolving in a given fixed background metric,\footnote{For simplicity we will only discuss scalar fields.} and the first-order perturbations of the metric and matter fields. In the former one usually neglects the effect of the field in the background geometry whereas the first-order metric perturbations are considered in the latter. Both systems become bad approximations when the fields or perturbations generate an energy-momentum tensor comparable with the background one.

We use the language of the Adiabatic Invariants of classical mechanics to obtain the third result. Requiring that the particle number density in the Ultraviolet (UV) limit depends only on the adiabatic invariants, we show that this leads naturally to the ``positive frequency'' requirement for the basis functions. When imposed at higher orders in the adiabatic expansion, the same requirement leads to the usual adiabatic vacuum (or adiabatic conditions as discussed in~\cite{Agullo2013,Agullo2015}). The link between the number of particles and the adiabatic invariant is known since the initial paper of~\citet{Parker1968}. However, this treatment is based on the Wentzel-Kramers-Brillouin (WKB) approximation (see for example~\cite{Fulling1989,Parker2012}). In this work, we follow the approach via Action Angle (AA) variables. This method has the advantage of separating the degrees of freedom in adiabatic (action) and fast (angle) variables. It is worth noting that these two variables are canonical and satisfy the Hamilton equations of motion.\footnote{In the WKB approach the direct connection with the Hamilton equations is lost and one is left to deal with an ad-hoc second order equation in a specific parametrization.} Hence, in this language, the adiabatic vacuum is described in the representations where the basis functions depend only on adiabatic variables up to a given order.

Finally, in the fourth result we explore the time- and space-dependent LCT and show that, for each adiabatic order, there is an exact canonical transformation connecting the original canonical variables with a new set of canonical variables adapted to that adiabatic order. We also show that these canonical transformations can be interpreted as smearing of the field by a classical function.

To obtain the first result, we need to perform the quantization of a free field in an arbitrary set of canonical variables. For this we follow~\citet{Wald1994}, where the formal approach appropriate for quantizing free fields is developed without assuming any special form for the canonical variables. In Sec.~\ref{sec:quant:gen} we make a brief review on the quantization method of free scalar fields in an arbitrary setting. We compare the method with the usual textbook approach by applying to the case of a free scalar field in a Friedmann--Lema\^{i}tre--Robertson--Walker (FLRW) background metric. Following, in Sec.~\ref{sec:rep} we also review the choice of representation in this formalism while comparing with the two other commonly used forms, namely, through the two point function or a complex structure. We also discuss the imposition of unitary evolution and its consequences in different representations. 

With these results in hand, we need to develop an asymptotic analysis of the solutions without imposing any initial conditions or choice of canonical variables, while imposing the normalization of the states. This is achieved using the AA variables and re-parameterizing the system such that the states are automatically normalized, independently of the initial conditions or canonical variables. In Sec.~\ref{sec:time:evol} we describe the asymptotic analysis in terms of the adiabatic invariants and obtain under which conditions the Hamiltonian allows the unitary evolution. Finally, in Sec.~\ref{sec:LCT} we study the group of time-dependent LCT and obtain the canonical variables in which the system has an unitary evolution. Considering a time- and space-dependent LCT, we show a recurrent series of canonical transformations relating each order of the adiabatic expansion with a particular exact LCT. In Sec.~\ref{sec:conclusions} we close this work presenting our conclusions.

\section{Quantization in arbitrary variables}
\label{sec:quant:gen}

The most common example of field quantization in a curved manifold is that of a massive free scalar field with a canonical kinetic term. In this context, the action is
\begin{equation}
S_\phi = -\frac{1}{2}\int\fint\left(g^{\mu\nu}\del_\mu\phi\del_\nu\phi + \mu^2\phi^2\right),
\end{equation}
where $\mu$ is the mass of the field $\phi$. Performing the $3+1$ splitting, the action is given by
\begin{equation}
S_\phi = \frac{1}{2}\int\ddx\ddt\;a^3\left(\dot{\phi}^2 + \frac{\phi\tilde{\scd}^2\phi}{a^2} - \mu^2\phi^2\right).
\end{equation}
The definitions and notation of the objects used in this work can be found in Appendix~\ref{app:def}.
We can also write this action making explicit the canonical structure. That is, the usual Legendre transform gives the momentum $\pmf = a^3\dot{\mf}$ and, thus,
\begin{equation}\label{eq:action:old}
\begin{split}
S_{\mf,\pmf} &= \int\ddx\ddt\left(\frac{\pmf\dot{\mf} - \mf\dpmf}{2} - \gH\right), \\
\gH &= \left(\frac{\pmf^2}{2a^3} - \frac{a\phi\tilde{\scd}^2\phi}{2} + \frac{a^3\mu^2\phi^2}{2}\right),
\end{split}
\end{equation}
where we wrote $\pmf\dot{\mf}$ as $(\pmf\dot{\mf} - \dpmf\mf)/2$ by removing a total derivative term.  

In this work we investigate the quantization of the fields using different canonical variables; it turns out to be useful to introduce a symplectic formalism for that purpose. We define the phase vector field $\chi_a$ and the symmetric Hamiltonian tensor $\gH^{ab}$, respectively, as 
\begin{equation}
\chi_a \doteq (\mf,\;\pmf),\qquad\gH(\chi) = \frac{1}{2}\chi_a\gH^{ab}\chi_b.
\end{equation}
Then,\footnote{In this work we use the symbol $\doteq$ to define components of vectors and tensors, and $\equiv$ for general definitions.} it is easy to see that the action is written as
\begin{equation}\label{eq:action:new}
\begin{split}
S_{\mf,\pmf} &= \frac{1}{2}\int\ddx\ddt\left(\ci\chi_a\SM^{ab}\dot{\chi}_b - \chi_a\gH^{ab}\chi_b\right),
\end{split}
\end{equation}
where the symplectic matrix and its inverse are defined by 
\begin{equation}\label{eq:def:SM}
\SM_{ab} \doteq \ci\left( \begin{array}{cc}
0 & 1 \\
-1 & 0 \end{array} \right),\quad \SM^{ab} \doteq \ci\left( \begin{array}{cc}
0 & 1 \\
-1 & 0 \end{array} \right),
\end{equation}
in which the imaginary unit $\ci$ is added for later convenience. For the system above, the Hamiltonian tensor is diagonal and its components are expressed by 
\begin{equation}\label{eq:H:KG}
\gH^{ab} \doteq \left( \begin{array}{cc}
-a\tilde{\scd}^2 + a^3\mu^2 & 0 \\
0 & \frac{1}{a^3} \end{array} \right).
\end{equation}
The action written in Eq.~\eqref{eq:action:new} is the same action of Eq.~\eqref{eq:action:old} but written in terms of the symplectic form and the Hamiltonian tensor. Given two phase vectors $\chi_a$ and $\sigma_a$ at the same time slice, the product $\chi_a\SM^{ab}\sigma_b$ is invariant under LCT. Using this formalism one can express the relevant quantities and operators in terms of phase space vectors. Consequently, those quantities written in terms of that product will be automatically recognized as invariant under LCT.

To relate this mathematical structure with quantization, we first note that the Poisson bracket of two field functionals $F_1$ and $F_2$ is given by
\begin{equation}
\Poisson{F_1}{F_2} = -\ci\int\limits_\ST\ddx\myfrac{\delta F_1}{\delta\chi_a(x)}\SM_{ab}\myfrac{\delta F_2}{\delta\chi_b(x)},
\end{equation}
where $\ST$ represents one particular spatial hypersurface of the $3+1$ splitting. Using the definitions above, it is easy to see that
\begin{equation}
\begin{split}
\Poisson{\chi_a(x_1)}{\chi_b(x_2)} &= -\ci\SM_{ab}\dirac{3}{x_1-x_2},
\end{split}
\end{equation}
where, i.e., for the scalar field case ($a=1$ and $b=2$), it reduces to the familiar expression, 
\begin{equation}
\Poisson{\mf(x_1)}{\Pi_{\mf}(x_2)} = \dirac{3}{x_1-x_2},
\end{equation}
as expected. Applying the canonical quantization rules, we promote the fields to Hermitian operators, which we denote with a hat, e.g., $\hat{\mf}$ and $\hat{\Pi}_{\mf}$ are the field operators related to the classical variables $\mf$ and $\Pi_{\mf}$, respectively. Therefore, following the canonical quantization, we have the equal-time commutation relations
\begin{align}
\Comm{\hat{\mf}(x_1)}{\hat{\Pi}_{\mf}(x_2)} &= \ci\dirac{3}{x_1-x_2}, \\
\Comm{\hat{\mf}(x_1)}{\hat{\mf}(x_2)} &= 0 = \Comm{\hat{\Pi}_{\mf}(x_1)}{\hat{\Pi}_{\mf}(x_2)},
\end{align}
which can be combined as a single expression
\begin{equation}\label{eq:def:comm:chi1:chi2}
\Comm{\mvo_a(x_1)}{\mvo_b(x_2)} = \SM_{ab}\dirac{3}{x_1 - x_2},
\end{equation}
where we are setting $\hbar = 1$. 

The canonical quantization defines the operators algebra, however, it does not provide a way to build the representations of this algebra. In QM, since we have a finite number of degrees of freedom, all representations are unitary equivalent (Stone--von Neumann theorem~\cite{Reed1979}). Therefore, in this case, the lack of a natural procedure for building the representations is irrelevant, being any representation physically equivalent to any other. On the other hand, for field quantization, we have an infinite number of degrees of freedom and the Stone--von Neumann theorem does not apply. This amounts to saying that, for fields, the canonical quantization no longer gives a complete description of the quantum system and one must complete it by choosing a representation or class of unitary equivalent representations.

The general procedure to obtain a representation starts by defining the product between two solutions of the equations of motion. These equations are derived from the action in Eqs.~\eqref{eq:action:old} [or equivalently Eq.~\eqref{eq:action:new}]. Given two solutions $u_1$ and $u_2$, the usual product can be defined as
\begin{equation}\label{eq:prod:klein}
\Prod{u_1}{u_2} = \ci\int\limits_\ST\dd^3x a^3 \left(u_1^*\dot{u}_2 - u_2\dot{u}_1^*\right).
\end{equation}
Assuming that we can find a complete set of solutions, in the sense that, 
\begin{equation}\label{eq:def:complete}
\Prod{u_{\HFi_1}}{u_{\HFi_2}} = \dirac{3}{\HFi_1-\HFi_2}, \qquad \Prod{u^*_{\HFi_1}}{u_{\HFi_2}} = 0,
\end{equation}
where $\HFi$ represents all the necessary indices to label all solutions,\footnote{For example, for flat spatial sections we can decompose the function in Fourier space and, in this case, $\HFi$ would be the Fourier mode vector. } we can then define the field operator as 
\begin{equation}\label{eq:mf:dec}
\hat{\mf} = \int\dd^3{}\HFi\left(u_{\HFi}\AO_{\HFi}+u_{\HFi}^*\AO_{\HFi}^\dagger\right),
\end{equation}
where $\int\dd^3\HFi$ represent all necessary sums and integrals over the indices $\HFi$. 

In the usual procedure, as in \cite{Birrell1982} for example, a set of annihilation and creation operators, respectively, $\AO_{\HFi}$ and $\AO_{\HFi}^\dagger$ is defined satisfying the usual commutation relations, i.e., 
\begin{equation}\label{eq:comm:a:ad}
\Comm{\AO_{\HFi_1}}{\AO^\dagger_{\HFi_2}} = \dirac{3}{\HFi_1-\HFi_2}\qquad \Comm{\AO_{\HFi_1}}{\AO_{\HFi_2}} = 0.
\end{equation}
From these operators one defines a Fock space, i.e., a vacuum such that $\AO_{\HFi}\ket{0} = 0$ for all indices $\HFi$, and all other states are defined by applying a finite number of operator $\AO^\dagger_{\HFi}$ with different indices $\HFi$ in $\ket{0}$. Hence, from the commutation relations given by Eq.~\eqref{eq:comm:a:ad} and the properties of the basis functions $u_{\HFi}$, one can show that the field operators defined by Eq.~\eqref{eq:mf:dec} will satisfy the correct commutation relations.

In short, in the procedure described in the paragraph above, one first defines a product between solutions. Then, given a complete set of solutions, one writes the field operator in terms of creation and annihilation operators with their commutation relations. Finally, using these results, one obtains the canonical commutation relations for the field operator. 

We find more convenient to invert this procedure and define the annihilation and creation operators from the canonical commutation of the field operators. Without assuming any decomposition for the field operator, we can calculate the products 
\begin{equation}
\Prod{u_{\HFi}}{\hat{\mf}}, \qquad \Prod{u^*_{\HFi}}{\hat{\mf}},
\end{equation}
which are also  operators. Therefore, we can compute the commutator between them using the canonical commutation relations between the field and its time derivative, i.e.,
\begin{equation}
\begin{split}
\Comm{\Prod{u_{\HFi_1}}{\hat{\mf}}}{\Prod{u_{\HFi_2}}{\hat{\mf}}} &= 0, \\ 
\Comm{\Prod{u^*_{\HFi_1}}{\hat{\mf}}}{\Prod{u^*_{\HFi_2}}{\hat{\mf}}} &= 0,
\end{split}
\end{equation}
result of the property $\Prod{u^*_{\HFi_1}}{u_{\HFi_2}} = 0$ and
\begin{equation}
\Comm{\Prod{u_{\HFi_1}}{\hat{\mf}}}{\Prod{u^*_{\HFi_2}}{\hat{\mf}}} = -\dirac{3}{\HFi_1-\HFi_2},
\end{equation}
which is a consequence of the other product of the basis functions $\Prod{u_{\HFi_1}}{u_{\HFi_2}} = \dirac{3}{\HFi_1-\HFi_2}$.\footnote{The minus sign in the expression above appears since $$\Prod{u^*_{\HFi_1}}{u^*_{\HFi_2}} = -\dirac{3}{\HFi_1-\HFi_2}.$$} Thus, it is natural to identify the products above as the annihilation and creation operators associated with the set of solutions $\{u_{\HFi}\}$, i.e., 
\begin{equation}
\AO_{\HFi} \equiv \Prod{u_{\HFi}}{\hat{\mf}}, \qquad \AO^\dagger_{\HFi} \equiv -\Prod{u^*_{\HFi}}{\hat{\mf}}.
\end{equation}
Finally, the completeness of the basis $\{u_{\HFi}\}$ can be expressed by the identity operation
\begin{equation}\label{eq:def:proj}
\int\dd^3\HFi\left[u_{\HFi}\Prod{u_{\HFi}}{f} - u^*_{\HFi}\Prod{u^*_{\HFi}}{f}\right] = f,
\end{equation}
for any function $f$. To see this, note that a function $f$ can be written as 
\begin{equation}
f(x) = \int\dd^3\HFi\left[c_{\HFi} u_{\HFi}(x) + d_{\HFi} u^*_{\HFi}(x)\right],
\end{equation}
for a fixed set of coefficients $c_{\HFi}$ and $d_{\HFi}$, since the basis is by assumption complete. Using the properties in Eq.~\eqref{eq:def:complete} it is easy to see that the operator in Eq.~\eqref{eq:def:proj} is just the identity operator. Applying the identity to the field operator and using our definitions of annihilation and creation operators, we recover Eq.~\eqref{eq:mf:dec}.

Both described procedures are constructed in terms of the set of solutions $\{u_{\HFi}\}$ of the equations of motion. It is however rarely the case where there is an analytic set of solutions. Then, to be able to explore the space of possible representations, we re-express the problem in phase space. This is possible because there is a natural one-to-one mapping between the solutions and their values in the phase space at a specific time slice.\footnote{This applies as long as the equations of motion have unique solutions.} Now, instead of having a set of solutions $\{u_{\HFi}(t,x)\}$, we have another set $\{\BF{\HFi,a}(t_0,x)\}$, where 
\begin{equation}
\BF{\HFi,a}(t_0,x) \doteq (u_{\HFi}(t_0,x),\Pi_{u_{\HFi}}(t_0,x)),
\end{equation}
are functions of the space slice labeled by $t_0$ representing the initial conditions for each mode. These later two functions are arbitrary and can be chosen without any knowledge of the solutions of the equations of motion.

We can express the product defined in Eq.~\eqref{eq:prod:klein} as
\begin{equation}
\Prod{\BF{\HFi_1}}{\BF{\HFi_2}} = \ci\int\limits_{\ST(t_0)}\dd^3x \left(u_{\HFi_1}^*\Pi_{u_{\HFi_2}} - u_{\HFi_2}\Pi^*_{u_{\HFi_1}}\right),
\end{equation}
where we omitted the time slice label and spatial position for simplicity, and used the momentum definition $\Pi_{u_1} = a^3\dot{u}_1$ provided by the action in Eq.~\eqref{eq:action:old}. Using the symplectic matrix, we can write the product in a compact form as
\begin{equation}\label{eq:prod:BF1:BF2}
\Prod{\BF{\HFi_1}}{\BF{\HFi_2}} = \int\limits_{\ST(t_0)}\dd^3x \BF{\HFi_1,a}^*\SM^{ab}\BF{\HFi_2,b}.
\end{equation}
The other expressions translate naturally, namely,
\begin{align}\label{eq:def:prod:BF}
&\Prod{\BF{\HFi_1}}{\BF{\HFi_2}} = \dirac{3}{\HFi_1-\HFi_2}, \qquad \Prod{\BF{\HFi_1}^*}{\BF{\HFi_2}} = 0, \\
&\int\dd^3\HFi\left[\BF{\HFi,a}\Prod{\BF{\HFi}}{f} - \BF{\HFi,a}^*\Prod{\BF{\HFi}^*}{f}\right] = f_a, \\ \label{eq:def:AO}
&\AO_{\HFi} = \Prod{\BF{\HFi}}{\mvo}, \qquad \AO^\dagger_{\HFi} = -\Prod{\BF{\HFi}^*}{\mvo}, \\ \label{eq:def:hatchi}
&\mvo_a = \int\dd^3{}\HFi\left(\BF{\HFi,a}\AO_{\HFi} + \BF{\HFi,a}^*\AO_{\HFi}^\dagger\right),
\end{align}
where new all quantities are expressed through phase space vectors.

Since we are working in phase space at a given spatial slice, it is reasonable to ask whether it will remain a complete basis at all times, given it is a complete basis on that slice. To address this question, we first note that the Hamilton equations, 
\begin{equation}
\dot{u}_{\HFi} = \myfrac{\del\gH}{\del\Pi_{u_{\HFi}}}, \quad \dot{\Pi}_{u_{\HFi}} = -\myfrac{\del\gH}{\del u_{\HFi}},
\end{equation}
can be conveniently expressed by
\begin{equation}\label{eq:motion:BF}
\ci\dBF{\HFi,a} = \SM_{ab}\myfrac{\del\gH}{\del\BF{\HFi,b}} = \SM_{ab}\gH^{bc}\BF{\HFi,c},
\end{equation}
where the first $=$ is general and the second applies to quadratic Hamiltonians.

Using two different initial conditions $\BF{\HFi_1,a}$ and $\BF{\HFi_2,a}$ at the same time $t_0$, we can use the equations of motion to build the solution for any time $t$, i.e., $\BF{\HFi_1,a}(t)$ and $\BF{\HFi_2,a}(t)$. Calculating the time derivative of the product Eq.~\eqref{eq:prod:BF1:BF2} on an arbitrary hypersurface labeled by $t$, we have that 
\begin{equation}\label{eq:cons:prod}
\begin{split}
\ci\lie_\n\Prod{\BF{\HFi_1}(t)}{\BF{\HFi_2}(t)} &=  0.
\end{split}
\end{equation}
Consequently, if $\BF{\HFi,a}$ initially forms a basis at $t=t_0$, it will remain a basis for all times.\footnote{The expression above also requires that $\SM_{ab}\gH^{bc}$ be self-adjoint with respect to the product.}

These results were derived having in mind the example of the Klein-Gordon (KG) field as shown in the action above. However, it extends for any system where the action is quadratic on the fields. For this reason, the product given in Eq.~\eqref{eq:def:prod:BF} is more general since it applies for any quadratic Hamiltonian system, while the product in Eq.~\eqref{eq:prod:klein} is specific for the KG field.

In brief, given a complete set of phase space functions $\BF{\HFi,a}$ defined on a spatial slice, we can decompose the quantum field in terms of creation and annihilation operators and, consequently, the representation as the Fock space defined by them, as summarized in Eqs.~(\ref{eq:def:prod:BF}--\ref{eq:def:hatchi}). The procedure is valid for any quadratic Hamiltonian system in canonical variables. It is not necessary to choose specific canonical variables for the field to apply this method.

\section{Representation}
\label{sec:rep}

The problem of finding a complete set of functions $\BF{\HFi,a}$ satisfying Eq.~\eqref{eq:def:prod:BF} can by simplified by focusing on functions of the form
\begin{equation}\label{eq:def:BF:split}
\BF{\HFi,a}(t_0) = \BTi_{\HFi,a}\HF{\HFi},
\end{equation}
where $\BTi_{\HFi,a}$ are arbitrary complex constants and $\HF{\HFi}$ are the Laplacian eigenfunctions defined in Eq.~\eqref{eq:laplace}. The product of two functions is given by
\begin{align}
\Prod{\BF{\HFi_1}}{\BF{\HFi_2}} &= \BTi_{\HFi_1a}^*\BTi_{\HFi_2b}\SM^{ab}\delta^3(\HFi_1-\HFi_2),\\
\Prod{\BF{\HFi_1}^*}{\BF{\HFi_2}} &= \BTi_{\HFi_1a}\BTi_{\HFi_2b}\SM^{ab}\delta^3(\HFi_1-\HFi_2) = 0.
\end{align}
The last equality results from the multiplication by a Dirac delta function that implies $\HFi_2 = \HFi_1$, which leads to zero since $\SM^{ab}$ is anti-symmetric. Thus, to have a normalized basis, we only need to impose 
\begin{equation}
\BTi_{a}^*\BTi_{b}\SM^{ab} = 1,
\end{equation}
where we omit the label $\HFi$ for simplicity.

This reduces the problem of finding a set of spatial functions $\BF{\HFi,a}$ that of finding an infinite set of bi-dimensional complex vectors $\BTi_{a}$, satisfying $\BTi_{a}^*\BTi_{b}\SM^{ab} = 1$. For this reason, it is useful to explore the properties of this vector space whereas we can use it to define arbitrary solutions for the basis functions. 

We can write the product between two vectors $\BRi_a$ and $\BSi_a$ as $\BRi_a^*\SM^{ab}\BSi_b.$ We will rise and lower indices using the matrix $\SM_{ab}$ and its inverse through $\BRi^a \equiv \SM^{ab}\BRi_b$. From this definition, we have the following properties 
\begin{equation}
\begin{split}
\BRi_a &= \SM_{ab}\BRi^b,\qquad \SM^{ab}\BRi_b = \BRi^a, \\ 
\BRi^{a*} &= \BRi_b^*\SM^{ba}, \qquad \BRi^{b*}\SM_{ba} = \BRi_a^*,\\
\BRi_a^*\BSi^a &= \BRi^{a*}\BSi_a.
\end{split}
\end{equation}

If a given vector has a positive norm, i.e., $\BRi_a^*\SM^{ab}\BRi_b > 0$, its complex conjugate has a negative norm $\BRi_a\SM^{ab}\BRi_b^* = -\BRi_b^*\SM^{ba}\BRi_a < 0$. Therefore, it is convenient to choose the basis such that the vector $\BRi_a$ always has a positive norm and its complex conjugate a negative one. Using a normalized vector, $\BRi^{a*}\BRi_a = 1$, we define the projectors
\begin{equation}\label{eq:project}
\begin{split}
\Prp_a{}^b \equiv \BRi_a\BRi^{b*}, \qquad \Prp_a{}^{b*} \equiv \BRi^*_a\BRi^b = \SM^{bc}\Prp_c{}^d\SM_{da}.
\end{split}
\end{equation}
It is easy to show that
\begin{equation}\label{eq:def:complet}
\begin{split}
\delta_a{}^b &= \Prp_a{}^b + \Prp_a{}^{b*}, \qquad \SM_{ab} = \Prp_{ab} - \Prp_{ab}^*,
\end{split}
\end{equation}
where
\begin{equation}
\begin{split}
\Prp_{ab} &= \Prp_a{}^c\SM_{cb} = \BRi_a\BRi_b^*.
\end{split}
\end{equation}
The expressions above are valid for any unitary vector. This implies that, for any basis, the imaginary part of the projector $\Prp_{ab}$ is always the symplectic matrix $\SM_{ab}$ (or, equivalently, the real part of $\Prp_a{}^b$ is always the identity). 

It is worth noting that the choice of the phase space vector $\BRi_a$ is arbitrary. There is not necessarily a connection with the field equations of motion. Some references~\cite{Fernandez-Mendez2012,Cortez2013,Gomar2012} always use the same fixed vector 
\begin{equation}\label{eq:def:bb}
\BRi_a \doteq \left(\frac{1}{\sqrt{2\EV{\HFi}}}, -\ci\sqrt{\frac{\EV{\HFi}}{2}}\right),
\end{equation}
independently of the field dynamics or time $t$. To relate this basis with the complex structure defined in these references, we can define the operator $-\ci\CS_a{}^b$ in terms of the projector $\Prp_a{}^b$, 
\begin{equation}\label{eq:def:CS}
\CS_a{}^b \equiv \left(\Prp_a{}^b - \Prp_a{}^{b*}\right), \qquad \CS_{ab} = \left(\Prp_{ab} + \Prp_{ab}^*\right),
\end{equation}
which provides a complex structure, i.e., 
\begin{equation}
(-\ci\CS_a{}^c)(-\ci\CS_c{}^b) = -\left(\Prp_a{}^b + \Prp_a{}^{b*}\right) = -\delta_a{}^b.
\end{equation}
Conversely, given a complex structure $\CS_a{}^b$, we can obtain a projector $\Prp_a{}^b$ defined by
\begin{equation}
\Prp_a{}^b \equiv \frac{\CS_a{}^b + \delta_a{}^b}{2}.
\end{equation}
Hence, the basis discussed in Eq.~\eqref{eq:def:bb} provides the following complex structure
\begin{equation}
-\ci\CS_a{}^b[\BRi] = \left( \begin{array}{cc}
0 & -\frac{1}{\EV{\HFi}} \\
\EV{\HFi} & 0 \end{array} \right),
\end{equation}
which is the same used in those works cited. We are not going to use this language when describing the quantization. Here we just showed how to compare the two formalisms. 

The choice of an arbitrary basis phase vector $\BRi_a$ a priori is not convenient as we will see later. It usually results in a vacuum definition in which the number of particles rapidly oscillates during the time evolution. On the other hand, we will show that, using the freedom in choosing the basis $\BRi_a$, there are vacuum definitions which are adiabatic, in the sense that the particle creation varies slowly when comparing to the frequency of the system, as expected.

\subsection{Two Point Functions}
\label{sec:twopf}

The physical interpretation of the phase space vectors described above is better understood by looking into the physical observables. For this reason, we can compute the two point function in the vacuum $\ket{0_{\BF{}}}$ of the representation defined by $\BF{\HFi,a}$,
\begin{equation}\label{eq:def:twopf}
\braketOP{0_{\BF{}}}{\mvo_a(x_1)\mvo_b(x_2)}{0_{\BF{}}} = \int\dd^3\HFi \BF{\HFi,a}(x_1)\BF{\HFi,b}^*(x_2),
\end{equation}
using Eq.~\eqref{eq:def:hatchi}. The anti-symmetric part of this two point function is simply the canonical commutator, as in Eq.~\eqref{eq:def:comm:chi1:chi2}. This can be written as 
\begin{equation}
\begin{split}
&\int\dd^3\HFi\left[\BF{\HFi,a}(x_1)\BF{\HFi,b}^*(x_2) - \BF{\HFi,b}(x_2)\BF{\HFi,a}^*(x_1)\right] = \\
&\int\dd^3\HFi\left(\BTi_a\BTi_b^* - \BTi_b\BTi_a^*\right)\HF{\HFi}(x_1)\HF{\HFi}(x_2) = \SM_{ab}\dirac{3}{x_1-x_2},
\end{split}
\end{equation}
by means of Eq.~\eqref{eq:def:BF:split}. Therefore, Eq.~\eqref{eq:def:complet} provides the connection between the canonical commutation and the basis $\BTi_a$, i.e., using the basis $\BTi_a$ to build the projector $\Prp_{ab}$ it is easy to see that
\begin{equation}
\Prp_{ab} = \BTi_a\BTi_b^*, \qquad \SM_{ab} = \BTi_a\BTi_b^* - \BTi_b\BTi_a^*.
\end{equation}
The product of the two operators in Eq.~\eqref{eq:def:twopf} can also be seen as the sum of the commutator and anti-commutator, hence,
\begin{equation}
\begin{split}
&\braketOP{0_{\BF{}}}{\mvo_a(x_1)\mvo_b(x_2)}{0_{\BF{}}} = \frac{1}{2}\SM_{ab}\dirac{3}{x_1 - x_2}\\
&+\frac12\int\dd^3\HFi \left[\BF{\HFi,a}(x_1)\BF{\HFi,b}^*(x_2) + \BF{\HFi,b}(x_2)\BF{\HFi,a}^*(x_1)\right],
\end{split}
\end{equation}
which, in terms of the splitting in Eq.~\eqref{eq:def:BF:split}, gives the last integral as
\begin{equation}
\begin{split}
&\frac{1}{2}\int\dd^3\HFi \left[\BF{\HFi,a}(x_1)\BF{\HFi,b}^*(x_2) + \BF{\HFi,b}(x_2)\BF{\HFi,a}^*(x_1)\right] = \\
&\frac{1}{2}\int\dd^3\HFi\left(\BTi_a\BTi_b^* + \BTi_b\BTi_a^*\right)\HF{\HFi}(x_1)\HF{\HFi}(x_2).
\end{split}
\end{equation} 
Putting these two equations together, we obtain
\begin{equation}\label{eq:twopf}
\begin{split}
&\braketOP{0_{\BF{}}}{\mvo_a(x_1)\mvo_b(x_2)}{0_{\BF{}}} = \frac{1}{2}\SM_{ab}\dirac{3}{x_1 - x_2}\\
&+\frac{1}{2}\int\dd^3\HFi\CS_{ab}\HF{\HFi}(x_1)\HF{\HFi}(x_2),
\end{split}
\end{equation}
where we used the operator defined in Eq.~\eqref{eq:def:CS}.

This leads us to conclude that the real part of the projector $\Prp_{ab}$ is what defines the two point function and, therefore, differentiates two representations. For example, if we choose $a=1$ and $b=1$ $\left( \hat{\chi}_1 = \hat{\phi} \right)$ we have 
\begin{equation}
\braketOP{0_{\BF{}}}{\hat{\mf}(x_1)\hat{\mf}(x_2)}{0_{\BF{}}} = \int\dd^3\HFi\vert\BTi_1\vert^2\HF{\HFi}(x_1)\HF{\HFi}(x_2).
\end{equation}
In other words, $\vert\BTi_1\vert^2$ is proportional to the power spectrum of the vacuum fluctuations in this representation.\footnote{The adimensional power spectrum is commonly defined as $$\Delta_{\hat{\mf}_{\HFi}} = \frac{\EV{\HFi}^3\vert\BTi_1\vert^2}{2\pi^2}.$$} Similarly, the power spectrum of the two point function of the field momentum, $\hat{\chi}_2 = \hat{\Pi}_{\phi}$,  is expressed by $\vert\BTi_2\vert^2$.

Before moving forward, it is convenient to introduce the following transformed operator
\begin{equation}\label{eq:trans:op}
\fsmvo_{\HFi,a} \equiv \int\limits_\ST\dd^3x \HF{\HFi}(x)\mvo_a(x),
\end{equation}
such that its commutator is 
\begin{equation}\label{eq:comm:fsmvo}
\Comm{\fsmvo_{\HFi_1,a}}{\fsmvo_{\HFi_2,b}} = \SM_{ab}\dirac{3}{\HFi_1-\HFi_2}.
\end{equation}
The two point function in the momentum space is thus
\begin{equation}
\begin{split}
&\braketOP{0_{\BF{}}}{\fsmvo_{\HFi_1,a}\fsmvo_{\HFi_2,b}}{0_{\BF{}}} = \left(\frac{\SM_{ab}+\CS_{ab}}{2}\right)\dirac{3}{\HFi_1 - \HFi_2}.
\end{split}
\end{equation}
From this, we can also study the two point function of the field and its momentum. Choosing $a=1$ and $b=2$ we obtain
\begin{equation}
\begin{split}
&\braketOP{0_{\BF{}}}{\fsmfo_{\HFi_1}\fspmfo{}_{\HFi_2}}{0_{\BF{}}} = \frac{1}{2}\left(\ci + \CS_{12}\right)\dirac{3}{\HFi_1 - \HFi_2}.
\end{split}
\end{equation}
In Ref.~\cite{Guth1985}, the balance between $\CS_{12}$ and $1$ ($=\hbar$ in ordinary unities) is argued to be related to the decoherence. 
Independently of the physical interpretation, $\CS_{12}$ provides a measure of how far from saturating are the Heisenberg uncertainty relations for the field and its momentum.

From the discussion above, we note that a basis can be interpreted through the analysis of the projector $\Prp_{ab}$ and, consequently, the complex structure $\CS_{ab}$. The independent components $\CS_{11}$, $\CS_{22}$ and $\CS_{12}$ can be interpreted in terms of the two points functions of the field, its momentum and their correlation, respectively. 

\subsection{Bogoliubov Coefficients}

Any product in the form $\Prod{\BRi\HF{\HFi}}{\mvo}$ can be expressed by
\begin{equation}
\Prod{\BRi\HF{\HFi}}{\mvo} = \BRi_a^*\SM^{ab}\fsmvo_b = \BRi^{a*}\fsmvo_a.
\end{equation}
Consequently, the annihilation and creation operators are 
\begin{equation}\label{eq:def:AO:fs}
\AO[\BRi]_{\HFi} = \BRi^{a*}\fsmvo_a, \qquad \AO^\dagger[\BRi]_{\HFi} = \BRi^{a}\fsmvo_a,
\end{equation}
which can be easily inverted using the projector in Eq.~\eqref{eq:def:complet} providing
\begin{equation}\label{eq:fsmvo:basis}
\fsmvo_a = \BRi_a\AO[\BRi]_{\HFi} + \BRi^*_a\AO^\dagger[\BRi]_{\HFi}.
\end{equation}
We will write the explicit dependency of the annihilation and creation operators on the basis $\BRi_a$ only when ambiguities can occur.

Each normalized phase space vector basis defines a representation for the quantization procedure. Two different basis, labeled respectively as $\BRi_a$ and $\BSi_a$, can lead to non-unitarily-equivalent representations, as we will see below. Assuming that $\BRi_a$ and $\BSi_a$ are two normalized phase space vectors and using the projectors in Eq.~\eqref{eq:def:complet}, we can write
\begin{equation}
\BRi_a = \alpha_{\BRi,\BSi}\BSi_a - \beta_{\BRi,\BSi}\BSi_a^*,
\end{equation}
where the products
\begin{equation}\label{eq:def:alpha:beta}
\alpha_{\BRi,\BSi} \equiv \BRi_a\BSi^{a*} = \BSi^*_a\SM^{ab}\BRi_b, \quad \beta_{\BRi,\BSi} \equiv -\BRi_a\BSi^a = \BSi_a\SM^{ab}\BRi_b,
\end{equation}
satisfy $\alpha_{\BRi,\BRi} = 1$ and $\beta_{\BRi,\BRi} = 0$. Using the projectors it is also easy to show that 
\begin{equation}\label{eq:alpha:beta:normal}
\vert\alpha_{\BRi,\BSi}\vert^2 - \vert\beta_{\BRi,\BSi}\vert^2 = 1.
\end{equation}
Then, the annihilation and creation operators defined by $\BRi_a$ can be written in terms of the equivalent operators defined by $\BSi_a$, namely,
\begin{equation}
\begin{split}
\AO_{\HFi}[\BRi] &= \alpha^*_{\BRi,\BSi}\AO_{\HFi}[\BSi] + \beta^*_{\BRi,\BSi}\AO^{\dagger}_{\HFi}[\BSi], \\
\AO_{\HFi}^{\dagger}[\BRi] &= \alpha_{\BRi,\BSi}\AO^{\dagger}_{\HFi}[\BSi] + \beta_{\BRi,\BSi}\AO_{\HFi}[\BSi].
\end{split}
\end{equation}
We note that, if $\beta_{\BRi,\BSi}$ vanishes (or equivalently $\Comm{\AO_{\HFi_1}[\BRi]}{\AO_{\HFi_2}[\BSi]} = 0$), both sets define the same vacuum. We express the vacuum defined by $\AO_{\HFi_1}[\BRi]$ as $\ket{0_\BRi}$, such that $\AO_{\HFi}[\BRi]\ket{0_\BRi} = 0$. Then, we define the number operator
\begin{equation}
N[\BRi] \equiv \int\dd^3\HFi\;\AO^{\dagger}_{\HFi}[\BRi]\AO_{\HFi}[\BRi]
\end{equation}
such that, if applied at $\ket{0_\BSi}$, it measures 
\begin{equation}\label{eq:unit:evol}
\braketOP{0_\BSi}{N[\BRi]}{0_\BSi} = \dirac{3}{0}\int\dd^3\HFi\left\vert\beta_{\BRi,\BSi}\right\vert^2,
\end{equation} 
where the $\dirac{3}{0}$ is the result of calculating the number of particles in a infinite spatial section. If the spatial section is compact, it would represent its volume. Thus, we formally define the particle number density operator as
\begin{equation}
n[\BRi] \equiv \frac{1}{\dirac{3}{0}}\int\dd^3\HFi\AO^{\dagger}_{\HFi}[\BRi]\AO_{\HFi}[\BRi],
\end{equation}
such that
\begin{equation}\label{eq:def:dens}
\braketOP{0_\BSi}{n[\BRi]}{0_\BSi} = \int\dd^3\HFi\left\vert\beta_{\BRi,\BSi}\right\vert^2.
\end{equation}
As discussed in Refs.~\cite{Fernandez-Mendez2012,Cortez2013,Gomar2012}, for example, any two representations are guaranteed to be unitarily equivalent if the integral in Eq.~\eqref{eq:def:dens} converges.\footnote{This discussion is much simpler when the spatial sections are compact. Here, we are always considering that we can use compact spatial section, such that the compatification scale is much larger than any scale involved in the problem.} In the following section we show that the convergence of the integral in Eq.~\eqref{eq:def:dens} is a necessary and sufficient condition for the two vacuums have non-zero internal product, i.e., $\braket{0_\BSi}{0_\BRi} \neq 0$.

\subsection{Wave Functional Representation}

It is useful to build the wave-functional representation for the Fock states. We are interested in the representation of the vacuum state for different choices of Fock representations. The operator $\fsmvo_{\HFi,a}$ is clearly an Hermitian operator and, as such, we can build its eigenvector basis. However, the two operators $\fsmvo_{\HFi,1}$ and  $\fsmvo_{\HFi,2}$ do not commute and, hence, we first build the eigenvector basis for the $\fsmvo_{\HFi,1}$ operator defining
\begin{equation}
\fsmfo_{\HFi} \equiv \fsmvo_{\HFi,1}, \qquad \fspmfo{}_{\HFi} \equiv \fsmvo_{\HFi,2}.
\end{equation}
We describe the eigenvector of the operators $\fsmfo_{\HFi}$ as $\ket{\fsmf}$, such that 
\begin{equation}
\fsmfo_{\HFi}\ket{\fsmf} = \fsmf_{\HFi}\ket{\fsmf},
\end{equation}
where $\fsmf_{\HFi}$ is a real function of $\HFi$. Given an arbitrary state $\ket{f}$, we write its projection on $\ket{\fsmf}$ as the wave functional 
\begin{equation}\label{eq:def:wf}
\wf{f}{\fsmf} \equiv \braket{\fsmf}{f}.
\end{equation}
In this scheme, the operators $\fsmfo_{\HFi}$ and $\fspmfo{}_{\HFi}$ are represented by 
\begin{equation}
\fsmfo_{\HFi} \wf{f}{\fsmf} = \fsmf_{\HFi} \wf{f}{\fsmf}, \qquad \fspmfo{}_{\HFi} \wf{f}{\fsmf} = -\ci\myfrac{\delta\wf{f}{\fsmf}}{\delta{}\fsmf_{\HFi}}.
\end{equation}
It is easy to see that it provides the right commutation relations [Eq.~\eqref{eq:comm:fsmvo}] for the fields.

Given a Fock representation defined by $\BTi_a$, the wave functional for its vacuum is $\wf{0_\BTi}{\fsmf}$. Projecting the two states $\bra{\fsmf}\fsmfo{}_{\HFi}$ and $\bra{\fsmf}\fspmfo{}_{\HFi}$ in the vacuum $\ket{0_\BTi}$, we obtain
\begin{equation}\label{eq:wf_vac}
\braketOP{\fsmf}{\fsmfo{}_{\HFi}}{0_\BTi} = \fsmf{}_{\HFi}\braket{\fsmf}{0_\BTi}.
\end{equation}
We use Eq.~\eqref{eq:fsmvo:basis} to write the operator $\fsmfo{}_{\HFi}$ in terms of the annihilation and creation operators, $\fsmfo{}_{\HFi} = \BTi_1\AO_{\HFi} + \BTi_1^*\AO^\dagger_{\HFi}$. Then, Eq.~\eqref{eq:wf_vac} reduces to
\begin{equation}
\BTi_1^*\braketOP{\fsmf}{\AO^\dagger_{\HFi}}{0_\BTi} = \fsmf{}_{\HFi}\wf{0_\BTi}{\fsmf}.
\end{equation}
Performing a similar calculation for $\fspmfo{}_{\HFi}$, we get
\begin{equation}\label{eq:def:rep:P}
\BTi_2^*\braketOP{\fsmf}{\AO^\dagger_{\HFi}}{0_\BTi} = -\ci\myfrac{\delta\wf{0_\BTi}{\fsmf}}{\delta\fsmf{}_{\HFi}},
\end{equation}
and, consequently, the wave functional satisfies
\begin{equation}
\myfrac{\delta\wf{0_\BTi}{\fsmf}}{\delta\fsmf{}_{\HFi}} = \frac{\ci\BTi_2^*}{\BTi_1^*}\fsmf{}_{\HFi}\wf{0_\BTi}{\fsmf} = \frac{\ci\CS_{12} - 1}{2\vert\BTi_1\vert^2}\fsmf{}_{\HFi}\wf{0_\BTi}{\fsmf},
\end{equation}
where we have used that $\BTi^{a*}\BTi_a = 1$. This functional equation has the simple Gaussian solution, i.e.,
\begin{equation}\label{eq:wf}
\wf{0_\BTi}{\fsmf} \sim \exp\left(-\int\dd^3\HFi\frac{1-\ci\CS_{12}}{4\vert\BTi_1\vert^2}\fsmf_{\HFi}^2\right).
\end{equation}

Note that we are using real basis functions $\HF{\HFi}$. Therefore, the operator $\fsmfo_{\HFi}$ is Hermitian and have real eigenvalues $\fsmf_{\HFi}$. However, in the literature one usually uses the Fourier transform of the field, resulting in a non-Hermitian operator and one has to deal with the real and imaginary parts of the eigenvalue. Alternatively, in the case where the spatial slice is flat, one can use the Hartley transform to obtain an Hermitian transform of the field, for example.\footnote{In the Hartley transform, we use $\sin(\bm{k}\cdot\bm{x}) + \cos(\bm{k}\cdot\bm{x})$ as the transform kernel, instead of $e^{-\ci\bm{k}\cdot\bm{x}}$. It has the advantage of being its own inverse and mapping real functions in real functions.} Given a real function $f(x)$, its Hartley transform is a real function 
\begin{equation}
\bar{f}^\text{H}(k) = \int\dd{}x\frac{\sin(kx) + \cos(kx)}{\sqrt{2\pi}}f(x).
\end{equation}
On the other hand, the Fourier transform results in a complex function, i.e., 
\begin{equation}
\bar{f}^\text{F}(k) = \int\dd{}x \frac{e^{-\ci kx}}{\sqrt{2\pi}}f(x).
\end{equation}
Then, it is easy to see that 
\begin{align}
\Re\left[\bar{f}^\text{F}(k)\right] &= \frac{\bar{f}^\text{H}(k) + \bar{f}^\text{H}(-k)}{2}, \\
\Im\left[\bar{f}^\text{F}(k)\right] &= \frac{\bar{f}^\text{H}(k) - \bar{f}^\text{H}(-k)}{2}.
\end{align}
Having these tools, we compare Eq.~\eqref{eq:wf} with the usual wave functional found in the literature (see, for example,~\cite[Eq.~(50)]{Martin2012}), namely,
\begin{equation}
\wf{0_\BTi}{\fsmf} \sim \exp\left(-2\int\limits_{\EV{\HFi}>0}\dd^3\HFi\Omega_{\HFi}\left[\Re\left(\fsmf^\text{F}_{\HFi}\right)^2 + \Im\left(\fsmf^\text{F}_{\HFi}\right)^2\right]\right),
\end{equation}
where the integration is done only in the half Fourier space. Using the relations between the Hartley and Fourier transforms, one can show that
\begin{equation}
\wf{0_\BTi}{\fsmf} \sim \exp\left(-\int\dd^3\HFi\Omega_{\HFi}(\fsmf^\text{H}_{\HFi})^2\right),
\end{equation}
and, thus, we identify 
\begin{equation}
\Omega_{\HFi} = \frac{1-\ci\CS_{12}}{4\vert\BTi_{\HFi}\vert^2} = -\frac{\ci}{2}\frac{\BTi^*_2}{\BTi^*_1}.
\end{equation}
Note also that, in the case where the momentum is simply the time derivative of the field, i.e., $\BTi_2 = \dot{\BTi}_1$, the expression for $\Omega_{\HFi}$ reduces to Eq.~(21) of~\cite{Martin2012}.

We find it simpler to work with real basis functions $\HF{\HFi}$ for two reasons: the wave functional depends on a single function (for a real field) avoiding working with half-Fourier spaces, and it is straightforward to generalize to other kinds of spatial sections.

We can now relate the unitary relation between two Fock spaces with the product between the vacuum of both spaces. We want to investigate what are the consequences of having a divergent integral in Eq.~\eqref{eq:def:dens}. Normalizing the wave functional [Eq.~\eqref{eq:wf}], we obtain the formal expression
\begin{equation}\label{eq:nwfT}
\wf{0_\BTi}{\fsmf} = \frac{e^{\int\dd^3\HFi\left[\ci\frac{\BTi_2^*}{2\BTi_1^*}\fsmf_{\HFi}^2-\frac{1}{4}\ln\left(\vert\BTi_1\vert^2\right)\right]}}{\sqrt{\prod_{\HFi}\sqrt{2\pi}}}.
\end{equation}
Analogously, given another field decomposition in terms of $\BRi_a$, the wave function representation of its vacuum is
\begin{equation}\label{eq:nwfR}
\wf{0_\BRi}{\fsmf} = \frac{e^{\int\dd^3\HFi\left[\ci\frac{\BRi_2^*}{2\BRi_1^*}\fsmf_{\HFi}^2-\frac{1}{4}\ln\left(\vert\BRi_1\vert^2\right)\right]}}{\sqrt{\prod_{\HFi}\sqrt{2\pi}}}.
\end{equation}
Thus, the internal product of these two states in the wave functional form is
\begin{equation}
\braket{0_\BRi}{0_\BTi} = \int\prod_{\HFi}\dd\fsmf_k\frac{e^{\int\dd^3\HFi\left[-\frac{1}{2}\frac{\BTi_a^*\BRi^a}{\BTi_1^*\BRi_1}\fsmf_{\HFi}^2-\frac{1}{4}\ln\left(\vert\BTi_1\vert^2\vert\BRi_1\vert^2\right)\right]}}{\prod_{\HFi}\sqrt{2\pi}}.
\end{equation}
Before evaluating the Gaussian integrals, we note that in the exponent we have $\alpha_{\BRi,\BTi} \equiv \BRi^a\BTi_{a*}$ [Eq.~\eqref{eq:def:alpha:beta}] and the factor that multiplies $\fsmf_{\HFi}^2$ can be expressed by 
\begin{equation}
\frac{\BTi_a^*\BRi^a}{\BTi_1^*\BRi_1} = \frac{\vert\alpha_{\BRi,\BTi}\vert}{\vert\BTi_1\vert\vert\BRi_1\vert}e^{\ci\varsigma},
\end{equation}
where $\varsigma$ is the combination of the phases of all three complex functions. In terms of these quantities the internal product is
\begin{equation}\label{eq:prod:v:v}
\braket{0_\BRi}{0_\BTi} = \exp{\int\dd^3\HFi\left[-\frac{\ci}{2}\varsigma-\frac{1}{4}\ln\left(\vert\alpha_{\BRi,\BTi}\vert^2\right)\right]}.
\end{equation}
This equation shows that the absolute value of the product is a function of the integral 
\begin{equation}
l_{\BTi,\BRi} \equiv \int\dd^3\HFi\ln\left(1 + \vert\beta_{\BRi,\BTi}\vert^2\right),
\end{equation}
where we used Eq.~\eqref{eq:alpha:beta:normal} to write $\alpha_{\BRi,\BTi}$ in terms of $\beta_{\BRi,\BTi}$. Clearly, if $l_{\BTi,\BRi}$ diverges, then the product between the two vacuums is zero.

We can show that, if the density of particles of $\BRi$ measured in the vacuum of $\BTi$ is finite, i.e., the two representations are unitarily related, then the product between the two vacuums is different from zero. Since the argument of the logarithm is always larger than one, the integrand will satisfy $\ln\left(1 + \vert\beta_{\BRi,\BTi}\vert^2\right) \geq 0$. Using the fact that $1+x^2 < \exp(x^2)$ or equivalently $\ln(1+x^2) < x^2$, it is clear that 
\begin{equation}
\int\dd^3\HFi\ln\left(1 + \vert\beta_{\BRi,\BTi}\vert^2\right) < \braketOP{0_\BTi}{n[\BRi]}{0_\BTi} = \int\dd^3\HFi\vert\beta_{\BRi,\BTi}\vert^2.
\end{equation}
Therefore, if $\braketOP{0_\BTi}{n[\BRi]}{0_\BTi} < \infty$, then $l_{\BTi,\BRi} < \infty$. This is equivalent to saying that if $l_{\BTi,\BRi}$ diverges then $\braketOP{0_\BTi}{n[\BRi]}{0_\BTi}$ diverges, i.e., if the two vacuums are orthogonal, then the number density of particles of one representation measured in the other is infinite.

In order to close the argument, we observe that a necessary condition for $l_{\BTi,\BRi}$ to converge is 
\begin{equation}
\lim_{\EV{\HFi}\rightarrow \infty} \EV{\HFi}^3\ln\left(1 + \vert\beta_{\BRi,\BTi}\vert^2\right) = 0,\quad \Rightarrow \quad \lim_{\EV{\HFi}\rightarrow \infty}\vert\beta_{\BRi,\BTi}\vert^2 = 0,
\end{equation}
where we are considering $\beta_{\BRi,\BTi}$ for which the limit exists. From here on, we will refer to the large $\EV{\HFi}$ limit as the UV limit. Therefore, this implies on 
\begin{equation}
\lim_{\EV{\HFi}\rightarrow \infty} \frac{\vert\beta_{\BRi,\BTi}\vert^2}{\ln\left(1 + \vert\beta_{\BRi,\BTi}\vert^2\right)} = 1,
\end{equation}
and consequently, using the limit comparison test, $\braketOP{0_\BTi}{n[\BRi]}{0_\BTi}$ converges if and only if $l_{\BTi,\BRi}$ converges.

\section{Time Evolution}
\label{sec:time:evol}

Until this point the discussion took place on a given spatial time slice. To extend to different times, we first note that, the splitting given in Eq.~\eqref{eq:def:BF:split} is really useful when the equations of motion are separable in the time variable. For example, when dealing with free fields or linear perturbations in a Friedmann background. When this happens, we can write the equation of motion [Eq.~\eqref{eq:motion:BF}] as
\begin{equation}
\ci\dBT_a = \SM_{ab}\gH^{bc}\BT_c,
\end{equation} 
where all Laplacian operators appearing in $\gH^{bc}$ must be substituted by $-\EV{\HFi}^2$, so that $\BT_a$ is a function of the time only.

To relate this function with the choice of representation $\BTi_a$, we impose that $\BT_a(t_0) = \BTi_a$, where $t_0$ labels the initial spatial slice.\footnote{The functions $\BT_a$ are always considered functions of $t$ and we will write the time dependency only to avoid ambiguities.} Since the product used to define the annihilation and creation operators is conserved [Eq.~\eqref{eq:cons:prod}], the operators defined in Eq.~\eqref{eq:def:AO} will be constant, i.e., they will represent these operators in the Schr\"{o}dinger picture. For example, the annihilation operator can be calculated at any time 
\begin{equation}
\AO_{\HFi} = \BT_{a}^*\fsmvo^a(t) = \BTi_{a}^*\fsmvo^a(t_0).
\end{equation}

The Heisenberg representation is defined through an evolution operator $\hat{E}(t, t_0)$ (that will not be necessarily unitary) such that
\begin{equation}\label{eq:def:AO:heisenberg}
\begin{split}
\AOh_{\HFi}(t) = \hat{E}^\dagger(t, t_0)\AO_{\HFi}\hat{E}(t, t_0) = \BTi_{a}^*\fsmvo^a(t),
\end{split}
\end{equation}
where we denote the Heisenberg annihilation (creation) operator as $\AOh_{\HFi}(t)$, which is always time-dependent. From here on we simplify the notation using only $\AOh_{\HFi}$ instead of $\AOh_{\HFi}(t)$. 

The annihilation and creation operators, in the Schr\"{o}dinger representation, can be written using two different forms,
\begin{align}
\AO_{\HFi} &= \BTi^{a*}\fsmvo_a(t_0), \qquad \AO_{\HFi}^\dagger = \BTi^{a}\fsmvo_a(t_0), \\
\AO_{\HFi}& = \BT^{a*}\fsmvo_a(t), \qquad \AO_{\HFi}^\dagger = \BT^{a}\fsmvo_a(t),
\end{align}
and in the Heisenberg's
\begin{equation}\label{eq:AOh:chi}
\AOh_{\HFi} = \BTi^{a*}\fsmvo_a(t), \qquad \AOh_{\HFi}^\dagger = \BTi^{a}\fsmvo_a(t).
\end{equation}
The two last set of equations can be inverted, as in Eq.~\eqref{eq:fsmvo:basis}, resulting in the two following expressions, respectively,
\begin{align}
\fsmvo_a(t) &= \BT_a\AO_{\HFi} + \BT^*_a\AO^\dagger_{\HFi}, \\
\fsmvo_a(t) &= \BTi_a\AOh_{\HFi} + \BTi^*_a\AOh^\dagger_{\HFi}.
\end{align}

The wave functional for a vacuum at late times $t$ is then 
\begin{equation}
\wf{0_{\BTi}}{\fsmf(t)} = \braket{\fsmf(t)}{0_\BTi} = \braket{\fsmf(t_0)}{0_\BT}.
\end{equation}
One can easily show that the wave functional at time $t$ is given by
\begin{equation}
\wf{0_\BT}{\fsmf} \sim \exp\left(-\int\dd^3\HFi\frac{1-\ci\CS_{12}[\BT]}{4\vert\BT_1\vert^2}\fsmf_{\HFi}^2\right).
\end{equation}

We can relate both representations using the Bogoliubov coefficients defined in Eq.~\eqref{eq:def:alpha:beta}, which we represent here as
\begin{equation}\label{eq:alpha:beta:t}
\alpha_{t,t_0} \equiv \alpha_{\BT,\BTi} = \BTi^*_a\BT^a, \quad \beta_{t,t_0} \equiv \beta_{\BT,\BTi} = \BTi_a\BT^a.
\end{equation}
Writing explicitly the annihilation and creation operators, we have
\begin{align}
\AOh_{\HFi} = \alpha_{t,t_0}\AO_{\HFi} - \beta^*_{t,t_0}\AO^\dagger_{\HFi},\quad& \AO_{\HFi} = \alpha^*_{t,t_0}\AOh_{\HFi} + \beta^*_{t,t_0}\AOh^\dagger_{\HFi},\\
\AOh^\dagger_{\HFi} = \alpha^*_{t,t_0}\AO^\dagger_{\HFi} - \beta_{t,t_0}\AO_{\HFi}, \quad&\AO^\dagger_{\HFi} = \alpha_{t,t_0}\AOh^\dagger_{\HFi} + \beta_{t,t_0}\AOh_{\HFi}.
\end{align}

Analyzing the behavior of $\beta_{t,t_0}$ as a function of $\HFi$, we can determine if the number density of particles in late times is finite [Eq.~\eqref{eq:def:dens}]. If $\braketOP{0_\BTi}{n[\BT]}{0_\BTi}$ is infinite after a finite amount of time $t$ has elapsed, then the vacuum defined by $\AOh_{\HFi}$ will be orthogonal to the initial vacuum [Eq.~\eqref{eq:prod:v:v}]. Since $\beta_{t,t_0}$ is defined as a product of two phase vectors on different time slices, it will not be invariant under time dependent LCT. 

It should be noted that the $\beta_{t,t_0}$ function usually does not represent the density of particles created between the times $t_0$ and $t$. It is the density of particles measured at $t$ {\bf if} the vacuum state of the observer at $t$ is still the same state $\ket{0_\BTi}$. To obtain the actual number of particles created, it is necessary to introduce a physically motivated vacuum at $t$ and measure $n[\BT]$ in this state. For example, one common option is to assume that an observer at rest in each hypersurface of a given foliation will perceive the adiabatic vacuum as its empty state. In this case, the number of particles created from an initial state defined at $t_0$ by $\BTi_a$ and measured at $t$ will be $\braketOP{0_{\text{adiab},t}}{n[\BT]}{0_{\text{adiab},t}}$.  For a detailed explanation about the choice of an adiabatic vacuum and its intrinsic ambiguity see~\cite{Chung2003}.

Nevertheless, the function $\beta_{t,t_0}$ is useful as a way to describe the ability of implementing the time evolution of the field operators using a unitary operator.

\subsection{Asymptotic behavior of the particle density function}
\label{sec:asymp:beta}

To determine if the integral defining the particle number density converges [Eq.~\eqref{eq:def:dens}], we need to study the behavior of $\beta_{t,t_0}$ for large $\EV{\HFi}$. The evolution of $\beta_{t,t_0}$ is defined in terms of the phase vector $\BT_a$ [Eq.~\eqref{eq:alpha:beta:t}] which satisfies the Hamilton equations of motion.

A general Hamiltonian for a free field in a homogeneous background is
\begin{equation}
\gH^{ab} \doteq \left( \begin{array}{cc}
\om\ofreq^2 & h \\
h & \frac{1}{\om} \end{array} \right),
\end{equation}
where the functions $\om$ and $\ofreq$ denote, respectively, the mass and frequency of the harmonic oscillator associated to each mode, and $h$ the coupling between the field and its momentum.

For a KG field [Eq.~\eqref{eq:H:KG}], the Hamiltonian decomposed in terms of $\HF{\HFi}$ is given by
\begin{equation}
\gH^{ab} \doteq \left[\begin{array}{cc}
a^3\left(\frac{\EV{\HFi}^2}{a^2} + \mu^2\right) & 0 \\
0 & \frac{1}{a^3} \end{array} \right],
\end{equation}
such that $\om = a^3$, 
\begin{equation}\label{eq:KG:freq}
\ofreq^2 = a^{-2}\EV{\HFi}^2 + \mu^2,
\end{equation}
and $h = 0$. In this case the UV limit of the frequency is 
\begin{equation}\label{eq:req:ofreq}
\lim_{\EV{\HFi}\rightarrow\infty}\ofreq\rightarrow \frac{\EV{\HFi}}{a}\rightarrow +\infty.
\end{equation}
In this work we consider accordingly the cases where $\ofreq$ has the asymptotic behavior of going to $+\infty$ when $\EV{\HFi}\rightarrow\infty$. This includes most of the physically motivated Hamiltonians. 

The curves defined by 
\begin{equation}\label{eq:E:curves}
\frac{1}{2}v_a\gH^{ab}v_b = \frac{1}{2}\om\ofreq^2q^2 + \frac{1}{2}\frac{p^2}{\om} + hpq = E,
\end{equation}
for a constant $E$ and real phase vector $v_a \doteq (q, p)$, are closed if $\ofreq^2 - h^2 > 0$. In the asymptotic limit, this is true if $\ofreq$ goes to infinity faster or at the same rate than $h$ while satisfying $\ofreq^2 - h^2 > 0$. If the Hamiltonian contains $h \neq 0$, it is possible to redefine the momentum using the canonical transformation 
\begin{equation}\label{eq:lct:rm:h}
q \rightarrow Q = q, \qquad p \rightarrow P = p + h\om q.
\end{equation}
Writing the Hamiltonian for these new variables, the frequency is redefined as 
\begin{equation}\label{eq:redef:freq}
\ofreq^2 \rightarrow \ofreq^2 - h^2 - \frac{\lie_\n(h\om)}{\om},
\end{equation}
and, consequently, 
\begin{equation}
\gH^{ab} \doteq \left( \begin{array}{cc}
\om\ofreq^2 & 0 \\
0 & \frac{1}{\om} \end{array} \right),
\end{equation}
where the new frequency is given by Eq.~\eqref{eq:redef:freq}. We define the Action-Angle (AA) canonical variables as 
\begin{equation}
I = \frac{v_a\gH^{ab}v_b}{2\ofreq}, \qquad \varphi = \arctan\left(\om\ofreq \frac{q}{p}\right),
\end{equation}
which satisfy the following equations of motions
\begin{equation}\label{eq:AA:eom}
\dot{I} = -2I\cos(2\varphi)\dot{\omfreq}, \quad \dot{\varphi} = \ofreq + \sin(2\varphi)\dot{\omfreq},
\end{equation}
where we defined the following function of only the background quantities
\begin{equation}\label{eq:def:zeta}
\omfreq \equiv \ln\left(\sqrt{\frac{\om\ofreq}{\om_0\ofreq_0}}\right),
\end{equation}
where $\om_0 \equiv \om(t_0)$, $\ofreq_0 \equiv \ofreq(t_0)$, the same convention (index ${}_0$ for a quantity evaluated at time $t_0$) applying in what follows. These quantities are related to the original variables $(q,p)$ by 
\begin{align}
q &= \sqrt{\frac{2I}{m\ofreq}}\sin(\varphi), \\ 
p &= \sqrt{2I\om\ofreq}\cos(\varphi).
\end{align}
These equations of motion have the advantage of making explicit the behavior of $I$ and $\varphi$ in the UV limit, i.e., the time derivative of the angle variable goes to infinity, while the derivative of the adiabatic variable has an oscillatory behavior which averages to zero when the frequency $\ofreq$ is much larger than $\dot{\omfreq}$.

Using two real solutions $(q_r, p_r)$ and $(q_i, p_i)$ we can write a general complex solution. It is convenient to shift the angle variable of the real solution by $\pi/2$, i.e.,
\begin{align}
q_r &= \sqrt{\frac{2I_r}{m\ofreq}}\cos(\varphi_r), \\ 
p_r &= -\sqrt{2I_r\om\ofreq}\sin(\varphi_r),
\end{align}
where the AA for the real part satisfy
\begin{equation}
\dot{I}_r = 2I_r\cos(2\varphi_r)\dot{\omfreq}, \quad \dot{\varphi}_r = \ofreq - \sin(2\varphi_r)\dot{\omfreq}.
\end{equation}
The complex solution can then be written as 
\begin{align}
\BT_1 &= \frac{q_r - \ci q_i}{2} = \frac{\sqrt{I_r}\cos(\varphi_r) - \ci \sqrt{I_i}\sin(\varphi_i)}{\sqrt{2m\ofreq}}, \\
\BT_2 &= \frac{p_r - \ci p_i}{2} = -\sqrt{\frac{\om\ofreq}{2}}\left[\sqrt{I_r}\sin(\varphi_r) + \ci \sqrt{I_i}\cos(\varphi_i)\right].
\end{align}
This complex solution has to satisfy the normalization condition $\BT_a^*\BT^a = 1$. The product of two solutions is always constant [Eq.~\eqref{eq:cons:prod}], therefore, if $\BT_a$ is initially normalized, it will remain so for all times. In terms of the AA variables, this restriction implies 
\begin{equation}\label{eq:def:norma}
I_rI_i = \frac{1}{\cos(\varphi_r - \varphi_i)^2}.
\end{equation}
Note that the variables $I_r$ and $I_i$ are the adiabatic invariants. Their derivatives average to zero, while $\varphi_r$ and $\varphi_i$ are the fast variables for which the time derivatives average to $\ofreq$. The state normalization above Eq.~\eqref{eq:def:norma} shows that the combination $\Delta\varphi \equiv \varphi_r - \varphi_i$ is also an adiabatic invariant, since it is a function of $I_r$ and $I_i$ only. This can also be seen by looking at the equation of motion for $\Delta\varphi$, i.e., 
\begin{equation}
\dot{\Delta\varphi} = 2\frac{\cos(2\bar{\varphi})}{\sqrt{I_rI_i}}\dot{\omfreq},
\end{equation}
where we define the mean angle variable 
\begin{equation}
\bar{\varphi} \equiv \frac{\varphi_r + \varphi_i}{2} + \frac{\pi}{4}.
\end{equation}
This shows that the time derivative of $\Delta\varphi$ also averages to zero.

From the equations of motion of $I_r$ and $I_i$, we can see that 
\begin{equation}
I_r = I_{r0}\exp\left[2\int\limits_{t_0}^t\dd{}t_1\cos(2\varphi_{r1})\dot{\omfreq}_1\right].
\end{equation}
Thus, if $I_{r0} > 0$, then $I_r > 0$ for all times. Using the normalization condition, we have only three real degrees of freedom for the initial conditions $I_{r0}$, $I_{i0}$ and $\bar{\varphi}_0$. We can choose $I_{r0} > 0$ and $I_{i0} > 0$ without loss of generality. Note also that the normalization condition [Eq.~\eqref{eq:def:norma}] imposes that $I_{r0}I_{i0} \geq 1$. The product $I_rI_i$ satisfies 
\begin{equation}
\lie_\n{\sqrt{I_rI_i}} = 2\sqrt{I_rI_i - 1}\cos(2\bar{\varphi})\dot{\omfreq},
\end{equation}
which we integrate obtaining
\begin{equation}
\begin{split}
&\sqrt{I_rI_i} = \\
&\cosh\left[\cosh^{-1}\left(\sqrt{I_{r0}I_{i0}}\right) + 2\int\limits_{t_0}^t\dd{}t_1\cos(2\bar{\varphi}_1)\dot{\omfreq}_1\right],
\end{split}
\end{equation}
and, therefore, $\sqrt{I_rI_i}$ satisfies $I_rI_i \geq 1$ for all times. The equation of motion for the mean angle variable is
\begin{equation}\label{eq:dot:barvarphi}
\dot{\bar{\varphi}} = \ofreq - \sqrt{\frac{I_rI_i - 1}{I_rI_i}}\sin(2\bar{\varphi}) \dot{\omfreq},
\end{equation}
and that for the ratio of the adiabatic invariants 
\begin{equation}
\lie_\n{\sqrt{\frac{I_r}{I_i}}} = 2\sqrt{\frac{I_r}{I_i}}\frac{\sin(2\bar{\varphi})}{\sqrt{I_rI_i}}\dot{\omfreq},
\end{equation}
which can be readily integrated as 
\begin{equation}
\sqrt{\frac{I_r}{I_i}} = \exp\left[\ln\left(\sqrt{\frac{I_{r0}}{I_{i0}}}\right) + 2\int\limits_{t_0}^t\dd{}t_1\frac{\sin(2\bar{\varphi}_1)}{\sqrt{I_rI_i}}\dot{\omfreq}_1\right].
\end{equation}
These integral equations naturally lead to the following parametrization 
\begin{equation}
I_r = \cosh(\epsilon)\exp(\gamma), \quad I_i = \cosh(\epsilon)\exp(-\gamma),
\end{equation}
where we introduced the functions
\begin{align}
\epsilon &\equiv \epsilon_0 + 2\int\limits_{t_0}^t\dd{}t_1\cos(2\bar{\varphi}_1)\dot{\omfreq}_1, \\
\gamma &\equiv \gamma_0 + 2\int\limits_{t_0}^t\dd{}t_1\frac{\sin(2\bar{\varphi}_1)}{\cosh(\epsilon)}\dot{\omfreq}_1,
\end{align}
where
\begin{equation}
\epsilon_0 = \cosh^{-1}\left(\sqrt{I_{r0}I_{i0}}\right), \quad \gamma_0 = \ln\left(\sqrt{\frac{I_{r0}}{I_{i0}}}\right).
\end{equation}
Using this parametrization, $\bar{\varphi}$ satisfies [see Eq.~\eqref{eq:dot:barvarphi}]
\begin{equation}\label{eq:dot:barvarphi2}
\dot{\bar{\varphi}} = \ofreq - \tanh(\epsilon)\sin(2\bar{\varphi})\dot{\omfreq}.
\end{equation}

We can now write $\BT_a$ directly in terms of the AA variables, 
\begin{align}\label{eq:BT1}
\BT_1 &= \frac{1}{2}\sqrt{\frac{1}{\om\ofreq}}\left(e^{-\ci\bar{\varphi}}F^c - \ci e^{\ci\bar{\varphi}}F^s\right), \\ \label{eq:BT2}
\BT_2 &=-\frac{\ci}{2}\sqrt{\om\ofreq}\left(e^{-\ci\bar{\varphi}}F^c + \ci e^{\ci\bar{\varphi}}F^s\right),
\end{align}
where we defined two functions which depend only on the adiabatic invariants,
\begin{align}\label{eq:def:Fc}
F^c &\equiv \cosh\left(\frac{\gamma+\epsilon}{2}\right) + \ci\cosh\left(\frac{\gamma-\epsilon}{2}\right),\\ \label{eq:def:Fs}
F^s &\equiv \sinh\left(\frac{\gamma-\epsilon}{2}\right) + \ci\sinh\left(\frac{\gamma+\epsilon}{2}\right).
\end{align}

In the UV limit $\EV{\HFi}\rightarrow\infty$, we can estimate the integrals defining $\epsilon$ and $\gamma$. The integral which defines $\delta\epsilon = \epsilon - \epsilon_0$ and $\delta\gamma \equiv \gamma - \gamma_0$ can be approximated as described in Appendix~\ref{app:intapprox}. From Eqs.~\eqref{eq:approx:de}, \eqref{eq:approx:dg} we conclude that the leading order approximation for $\delta\epsilon$ and $\delta\gamma$ depend on $\dot{\omfreq}/\ofreq$, and their higher order corrections depend on higher time derivatives and powers of $\ofreq^{-1}$, e.g., the second order terms will depend on terms of the form $$\frac{\dot{\omfreq}^2}{\ofreq^2}, \quad\text{and}\quad \frac{1}{\ofreq}\lie_\n\left(\frac{\dot{\omfreq}}{\ofreq}\right).$$ Therefore, we need to evaluate both $\dot{\omfreq}$ and $\ofreq$ to retrieve their spectral dependency.

The frequency of a KG field is given by Eq.~\eqref{eq:KG:freq}. We can generalize it writting  
\begin{equation}\label{eq:def:mv}
(\om\ofreq)^2 = \sum_{i=-2}^{\infty}f_{i}(t)\EV{\HFi}^{-i},
\end{equation}
where $f_i(t)$ are real functions of time only and $f_{-2}(t) > 0$.\footnote{The first term of the sum being $i=-2$ and the condition on $f_{-2}(t)$ are equivalent to the requirements on $\ofreq$ discussed below Eq.~\eqref{eq:req:ofreq}.} Using this expression we obtain 
\begin{equation}\label{eq:lim:dotmv}
\lim_{\EV{\HFi}\rightarrow\infty}\dot{\omfreq} = \frac{1}{4}\frac{\sum_{i=-2}^{\infty}\dot{f}_{i}(t)\EV{\HFi}^{-i}}{\sum_{i=-2}^{\infty}f_{i}(t)\EV{\HFi}^{-i}}.
\end{equation}
In general, this limit is of order $\OO{\EV{\HFi}^0}$. However, if $\dot{f}_{-2}(t) = 0$, then it is $\OO{\EV{\HFi}^{-1}}$ and, if both $\dot{f}_{-2}(t) = 0$ and $\dot{f}_{-1}(t) = 0$, then it is $\OO{\EV{\HFi}^{-2}}$ and so forth.\footnote{Note that the first term does not need to be proportional to $\EV{\HFi}^2$. If the largest power of the eigenvalue is $\EV{\HFi}^m$, then the same conclusions hold, e.g. $\dot{f}_{-m}(t) = 0$ implies $\dot{\omfreq} \propto \OO{\EV{\HFi}^{-1}}$, etc.}

This shows that if we do not assume anything about $\dot{\omfreq}$ other than Eq.~\eqref{eq:def:mv}, then this function is at least of order $\OO{\EV{\HFi}^0}$. Since the higher order terms in the approximation of $\delta\epsilon$ and $\delta\gamma$ contain an additional factor of $\ofreq^{-1}$, we conclude that these corrections are at least of order $\OO{\EV{\HFi}^{-1}}$ smaller than the first term. Hence, in the UV limit the first term is a good approximation for $\delta\epsilon$. On the other hand, the first term is also at least of order $\OO{\EV{\HFi}^{-1}}$ and, therefore, in this limit $\epsilon = \epsilon_0 + \OO{\EV{\HFi}^{-1}}$ and $\gamma = \gamma_0 + \OO{\EV{\HFi}^{-1}}$. In short, the asymptotic analysis in the UV limit above shows that the adiabatic invariants are constants in the limit of large $\EV{\HFi}$ and the time dependent corrections are all at least of order $\OO{\EV{\HFi}^{-1}}$. 

The function $\beta_{t,t_0}$ can be schematically written as 
\begin{equation}\label{eq:beta:AA}
\beta_{t,t_0} = \beta^{+}_{t,t_0}(\epsilon,\gamma,\omfreq)e^{\ci\bar{\varphi}} + \beta^{-}_{t,t_0}(\epsilon,\gamma,\omfreq)e^{-\ci\bar{\varphi}}.
\end{equation}
Since $\bar{\varphi}$ is a fast variable, the $\beta_{t,t_0}$ function oscillates rapidly, while the coefficients depend only on adiabatic invariants and $\omfreq$. We need to evaluate under which conditions $\vert\beta_{t,t_0}\vert$ satisfies the necessary condition to converge, i.e, $$\lim_{\EV{\HFi}\rightarrow \infty}\vert\beta_{t,t_0}\vert^2 = 0.$$ From Eq.~\eqref{eq:beta:AA} it is clear that while $\bar{\varphi}$ varies by $2\pi$ the other quantities remain constant. Then, the function $\beta_{t,t_0}$ will satisfy the required limit if and only if it does so for any value of $\bar{\varphi} \in (0,2\pi)$. This is equivalent to saying that both $\beta^{\pm}_{t,t_0}$ must go to zero in this limit. These variables are given by the following expressions
\begin{align}
\beta^-_{t,t_0} &\equiv \frac{1}{2}\left[F^c\sinh(\omfreq)e^{-\ci\bar{\varphi}_0}F^c_0-\ci F^c\cosh(\omfreq)e^{\ci\bar{\varphi}_0}F^s_0\right], \\
\beta^+_{t,t_0} &\equiv \frac{1}{2}\left[\ci F^s\cosh(\omfreq)e^{-\ci\bar{\varphi}_0}F^c_0 + F^s\sinh(\omfreq)e^{\ci\bar{\varphi}_0}F^s_0\right].
\end{align}
We can write the condition for both variables being zero as the matrix product
\begin{equation}\label{eq:det:beta}
\left(\begin{array}{cc}
F^c\sinh(\omfreq) & -\ci F^c\cosh(\omfreq) \\
\ci F^s\cosh(\omfreq) & F^s\sinh(\omfreq) \end{array}\right)\left(\begin{array}{c}
e^{-\ci\bar{\varphi}_0}F^c_0 \\
e^{\ci\bar{\varphi}_0}F^s_0 \end{array}\right) = \left(\begin{array}{c}
0 \\
0 \end{array}\right).
\end{equation} 
If this matrix is non-singular, then the equation above implies $F^c_0 = 0 = F^s_0$. However, a quick glance on Eqs.~\eqref{eq:def:Fc} and \eqref{eq:def:Fs} shows that this solution is impossible, $F^c$ has a minimum value different from zero. Thus, we must choose the initial conditions such that the matrix above is singular. Its determinant is given by
\begin{equation}
\left\vert\begin{array}{cc}
F^c\sinh(\omfreq) & -\ci F^c\cosh(\omfreq) \\
\ci F^s\cosh(\omfreq) & F^s\sinh(\omfreq) \end{array}\right\vert = \sinh(\epsilon) - \ci\cosh(\epsilon)\sinh(\gamma).
\end{equation}
Since $\delta\epsilon$ and $\delta\gamma$ already go to zero in this limit, this condition imposes that $\epsilon_0$ and $\gamma_0$ must also go to zero, i.e., the initial conditions must satisfy
\begin{equation}\label{eq:cond:e0g0}
\lim_{\EV{\HFi}\rightarrow \infty}\epsilon_0 = 0, \qquad \lim_{\EV{\HFi}\rightarrow \infty}\gamma_0 = 0,
\end{equation}
which translate into
\begin{equation}\label{eq:lim:beta:pm}
\begin{split}
\lim_{\EV{\HFi}\rightarrow \infty}\beta^{-}_{t,t_0} &= \frac{1}{2}\sinh(\omfreq)e^{-\ci\bar{\varphi}_0}, \\
\lim_{\EV{\HFi}\rightarrow \infty}\beta^{+}_{t,t_0} &= 0.
\end{split}
\end{equation}
Therefore, even after this choice of initial conditions, $\vert\beta_{t,t_0}\vert$ can only have the required limit provided the condition
\begin{equation}\label{eq:cond1:zeta}
\lim_{\EV{\HFi}\rightarrow \infty}\omfreq = \lim_{\EV{\HFi}\rightarrow\infty}\frac{1}{2}\ln\left(\frac{\om\ofreq}{\om_0\ofreq_0}\right) = 0
\end{equation}
on the background function holds. For the function $\om\ofreq$ in the form of Eq.~\eqref{eq:def:mv}, this requirement translates into
\begin{equation}\label{eq:zeta:expa}
\begin{split}
&\lim_{\EV{\HFi}\rightarrow \infty}\ln\left(\frac{\sum_{i=-2}^{\infty}f_{i}(t)\EV{\HFi}^{-i}}{\sum_{i=-2}^{\infty}f_{i}(t_0)\EV{\HFi}^{-i}}\right) = \ln\left(\frac{f_{-2}(t)}{f_{-2}(t_0)}\right) \\
&+ \left(\frac{f_{-1}(t)}{f_{-2}(t)}-\frac{f_{-1}(t_0)}{f_{-2}(t_0)}\right)\frac{1}{\EV{\HFi}} + \OO{\EV{\HFi}^{-2}}.
\end{split}
\end{equation}
Then, independently of any initial condition choice, the function $\beta_{t,t_0}$ will only have the required asymptotic behavior if $f_{-2}(t)$ is constant. Later we will show that changing the field representation through LCT, we can modify the function $\om\ofreq$ in order to obtain $f_{-2}(t)$ constant. This amounts to saying that the imposition of unitary evolution requires a particular choice of a family of canonical variables in which the above limit is achieved.

It is worth noting that the limit in Eq.~\eqref{eq:zeta:expa} relates the UV with the background variables on two time slices $t_0$ and $t$. It is a commonplace in QFT in curved background to study the UV limit in the asymptotic regions with $t_0$ and $t$ fixed (usually $t_0\rightarrow-\infty$ and $t\rightarrow\infty$) see for example Eq.~(2.24) in \cite{Agullo2015}. In this case, it is enough that the functions $f_{-2}$ and $f_{-1}$ have the same limit in both time coordinates. However, here we are interested in the UV limit between any two time slices $t_0$ and $t$. This provides stronger constraints on $f_{-2}$ and $f_{-1}$ which need to be constant instead of just coinciding on two fixed time slices only.

The necessary condition on $\omfreq$ [Eq.~\eqref{eq:cond1:zeta}] restricts the choice of canonical variables describing the field, as we will see explicitly later. However, the particle number density requires a stronger condition, $\lim_{\EV{\HFi}\rightarrow \infty}\EV{\HFi}^3\vert\beta_{t,t_0}\vert^2 = 0$, to converge. This condition puts constraints on how fast each element of $\beta_{t,t_0}$ must approach zero. Assuming that $\epsilon_0$ and $\gamma_0$ satisfy Eq.~\eqref{eq:cond:e0g0} and since $\delta\epsilon$ and $\delta\gamma$ also go to zero, we have that in the limit both $\epsilon$ and $\gamma$ go to zero. The unitary evolution already requires that $\omfreq$ goes to zero in the limit. Using these behaviors, we can expand every term of $\beta_{t,t_0}$ in powers of these functions. The lowest order approximations are
\begin{align}\label{eq:betam:fo}
\beta_{t,t_0}^- &\approx \frac{1}{2} \left[2 \ci \omfreq e^{-\ci \bar{\varphi}_0} + \left(\gamma_0  + \ci \epsilon_0\right)e^{\ci\bar{\varphi}_0}\right], \\ \label{eq:betap:fo}
\beta_{t,t_0}^+ &\approx -\frac{1}{2} e^{-\ci \bar{\varphi}_0} (\gamma +\ci\epsilon ).
\end{align}
Arranging the terms above, we conclude that for $\lim_{\EV{\HFi}\rightarrow \infty}\EV{\HFi}^3\vert\beta_{t,t_0}\vert^2 = 0$ we need 
\begin{equation}
\lim_{\EV{\HFi}\rightarrow \infty}\left\{\begin{array}{c}
\EV{\HFi}^{\frac{3}{2}}\omfreq = 0, \\
\EV{\HFi}^{\frac{3}{2}}\epsilon = 0, \\
\EV{\HFi}^{\frac{3}{2}}\gamma = 0.
\end{array}\right.
\end{equation}
Assuming that $\omfreq$ satisfies the first condition [Eq.~\eqref{eq:cond1:zeta}], we can use the expansion on Eq.~\eqref{eq:zeta:expa} to write 
\begin{equation}\label{eq:cond:omfreq}
\lim_{\EV{\HFi}\rightarrow \infty}\EV{\HFi}^{\frac{3}{2}}\omfreq = \frac{f_{-1}(t)-f_{-1}(t_0)}{f_{-2}(t_0)}\EV{\HFi}^{\frac12} + \OO{\EV{\HFi}^{-\frac12}}.
\end{equation}
Therefore, the particle number density integral will converge only if $f_{-1}(t)$ is also a constant. Consequently, the last requirement implies that 
\begin{equation}\label{eq:omfreq:lim}
\omfreq \propto \OO{\EV{\HFi}^{-2}}.
\end{equation}

The other variables $\epsilon_0$ and $\gamma_0$ depend on the choice of vacuum and can be adjusted freely in order to obtain a unitary evolution. Thus, we can always choose the initial conditions satisfying 
\begin{equation}
\lim_{\EV{\HFi}\rightarrow \infty}\EV{\HFi}^{\frac{3}{2}}\epsilon_0 = 0,\quad \lim_{\EV{\HFi}\rightarrow \infty}\EV{\HFi}^{\frac{3}{2}}\gamma_0 = 0.
\end{equation}
Finally, we must show that $\delta\epsilon$ and $\delta\gamma$ also go to zero at the required rate. From Eqs.~\eqref{eq:approx:de}, \eqref{eq:approx:dg}, we note that the leading terms are in the form $$\frac{1}{\ofreq}\times\dot{\omfreq}\propto \OO{\EV{\HFi}^{-1}} \times \OO{\EV{\HFi}^{-2}} \propto \OO{\EV{\HFi}^{-3}},$$ where we used the requirement on $\omfreq$ from \eqref{eq:omfreq:lim}. Therefore, the previous requirement is sufficient to make $\delta\epsilon$ and $\delta\gamma$ have the necessary limits.

We conclude that it is possible to choose an initial condition (a vacuum condition), where the time evolution is unitarily implemented, when the factor $\om\ofreq$ is such that $f_{-2}(t)$ and $f_{-1}(t)$ are constant. This, however, does not impose any constraint on the choice of initial conditions. In other words, this is a requirement on the functions appearing in the Hamiltonian and is independent of any choice of initial conditions. In Sec.~\ref{sec:LCT} we show how to use LCT to shape the functions in the Hamiltonian such that the unitary evolution is possible. 

Comparing again with the analysis in time asymptotic regions with $t_0$ and $t$ fixed, mentioned above. If the functions $f_{-2}$ and $f_{-1}$ coincide only in the asymptotic time limits $t_0$ and $t$, the main contribution to $\beta_{t,t_0}$ coming from $\omfreq$ will vanish [Eq.~\eqref{eq:betam:fo}]. However, the derivative of $\omfreq$ is present in $\delta\epsilon$ and $\delta\gamma$. Thus, if $f_{-2}$ and $f_{-1}$ are not constant, $\dot{\omfreq} \propto \OO{1}$. Consequently, the presence of such terms in $\beta_{t,t_0}$ [Eqs.~\eqref{eq:betam:fo} and \eqref{eq:betap:fo}] would result in an infinite particle number density.

\subsection{Vacuum stability}
\label{sec:vacuum:stab}

The particle number density for a general vacuum is described by the integral of Eq.~\eqref{eq:beta:AA}. Note that this quantity depends on the fast variable $\bar{\varphi}$ through oscillatory functions. This means that the particle number density fluctuates with frequency $\approx\ofreq$. We denote this behavior as an unstable vacuum for which the particle number for each mode oscillates between the minimum and maximum $\vert\beta_{t,t_0}\vert^2$ in each time interval $\approx 1/\ofreq$. 

We already know the dependency of $\delta\epsilon$, $\delta\gamma$ and $\omfreq$ on $\EV{\HFi}$, namely, $\omfreq$ goes slower to zero ($\propto \EV{\HFi}^{-2}$) while the other two are $\propto \EV{\HFi}^{-3}$. Therefore, we must choose $\epsilon_0$ and $\gamma_0$ in order to determine the behavior of $\beta_{t,t_0}$ in the large $\EV{\HFi}$ limit. If the initial conditions $\epsilon_0$ and $\gamma_0$ go slower than $\omfreq$ to zero, than the largest term in the expansion of $\beta_{t,t_0}$ is 
\begin{equation}
\beta_{t,t_0} \approx -\ci(\gamma_0+\ci\epsilon_0)\sin(\bar{\varphi} - \bar{\varphi}_0),
\end{equation}
which by our definition is an unstable vacuum, since the particle number density depends on the fast variable through the $\sin$ function. On the other hand, if we choose the initial conditions to vanish faster than $\omfreq$, we get another asymptotic behavior for the largest term, i.e., 
\begin{equation}
\beta_{t,t_0} \approx \ci\omfreq{}e^{-\ci(\bar{\varphi}-\bar{\varphi}_0)}.
\end{equation}
In this case the largest term of $\vert\beta_{t,t_0}\vert^2$ is non-oscillatory. Therefore, any initial condition satisfying 
\begin{equation}\label{eq:uv:init}
\lim_{\EV{\HFi}\rightarrow\infty}\frac{\epsilon_0}{\omfreq} = \lim_{\EV{\HFi}\rightarrow\infty}\frac{\gamma_0}{\omfreq} = 0,
\end{equation}
provides a vacuum which is stable at leading order in the UV limit. 

For the leading order approximation, we can use 
\begin{equation}
\epsilon = 0, \quad \gamma = 0 \quad \text{and}\quad \bar{\varphi} = \int\limits_{t_0}^t\dd{}t_1\ofreq_1 + \frac{\pi}{4},
\end{equation}
where $\ofreq_1 \equiv \ofreq(t_1)$, such that  the  Eqs.~\eqref{eq:BT1} and \eqref{eq:BT2} are
\begin{align}\label{eq:BT1:leading}
\BT_1 &= \frac{e^{-\ci\int\limits_{t_0}^t\dd{}t_1\ofreq_1}}{\sqrt{2\om\ofreq}} + \OO{\frac{\dot{\omfreq}}{\ofreq}}, \\ \label{eq:BT2:leading}
\BT_2 &= -\ci\sqrt{\frac{\om\ofreq}{2}}e^{-\ci\int\limits_{t_0}^t\dd{}t_1\ofreq_1} + \OO{\frac{\dot{\omfreq}}{\ofreq}}.
\end{align}
This approximation depends on $\EV{\HFi}$ through the oscillatory frequency $\ofreq$ in the exponent and the denominator $\sqrt{2\om\ofreq}$. In the UV limit it reduces to
\begin{equation}
\lim_{\EV{\HFi}\rightarrow\infty}\BT_1 = \frac{e^{-\ci\EV{\HFi}\int\limits_{t_0}^t\dd{}t_1\frac{1}{\om_1}}}{\sqrt{2\EV{\HFi}}}.
\end{equation}
Naturally, this leading order approximation coincides with the leading order of the WKB approximation (see~\cite{Fulling1989}, for example).

The next order terms will contain $\epsilon_0$, $\gamma_0$, $\delta\epsilon$ and $\delta\gamma$. The positive frequency part $\beta^{+}_{t,t_0}$ [Eq.~\eqref{eq:betap:fo}] contains a term of the form $\delta\gamma + \ci\delta\epsilon$. Using the asymptotic expansions given in Eqs.~\eqref{eq:approx:de} and \eqref{eq:approx:dg}, we obtain
\begin{equation}
\delta\gamma + \ci\delta\epsilon \approx -\left.e^{-2\ci\bar{\varphi}_1}\frac{\dot{\omfreq}_1}{\ofreq_1}\right\vert_{t_0}^{t}.
\end{equation}
Consequently, the term containing $\beta^+_{t,t_0}$ [Eq.~\eqref{eq:beta:AA}] at this order is given by
\begin{equation}
\begin{split}
e^{\ci\bar{\varphi}}\beta^+_{t,t_0} &\approx \frac{e^{-\ci(\bar{\varphi}+\bar{\varphi}_0)}}{2}\frac{\dot{\omfreq}}{\ofreq} \\
&-\frac{e^{\ci(\bar{\varphi}-\bar{\varphi}_0)}}{2}\left(\gamma_0 +\ci\epsilon_0 + e^{-2\ci\bar{\varphi}_0}\frac{\dot{\omfreq}_0}{\ofreq_0}\right).
\end{split}
\end{equation}
The first term only depends on the initial conditions through $\bar{\varphi}_0$ and has the same dependency on the fast variable of the main term, i.e.,
\begin{equation}
\begin{split}
\beta_{t,t_0} &\approx \left[\ci\omfreq{}e^{-\ci\bar{\varphi}_0} + \frac{e^{-\ci\bar{\varphi}_0}}{2}\frac{\dot{\omfreq}}{\ofreq} +  \left(\gamma_0  + \ci \epsilon_0\right)\frac{e^{\ci\bar{\varphi}_0}}{2}\right]e^{-\ci\bar{\varphi}} \\
&-\frac{e^{\ci(\bar{\varphi}-\bar{\varphi}_0)}}{2}\left(\gamma_0 +\ci\epsilon_0 + e^{-2\ci\bar{\varphi}_0}\frac{\dot{\omfreq}_0}{\ofreq_0}\right).
\end{split}
\end{equation}
The second line of the equation above oscillates with positive exponent on $\bar{\varphi}$ and, therefore, contributes with an oscillatory correction for the non-oscillatory first term. This shows that, when we calculate the higher order corrections for $\beta_{t,t_0}$, we find terms which contribute both in oscillatory and non-oscillatory manners. We denote this kind of vacuum as unstable at $\OO{\EV{\HFi}^{-3}}$, which is the order of the above correction. If we choose the initial conditions such that the second term is exactly zero, we obtain a choice of stable vacuum at $\OO{\EV{\HFi}^{-3}}$. This choice is simply
\begin{equation}\label{eq:first:stable}
\gamma_0 +\ci\epsilon_0 \approx - e^{-2\ci\bar{\varphi}_0}\frac{\dot{\omfreq}_0}{\ofreq_0}.
\end{equation}
We write the approximate equality since this condition fixes the initial condition up to $\OO{\EV{\HFi}^{-3}}$ terms. However, we need additional higher order terms to remove the high order instabilities, and this imposes higher order restrictions on the initial conditions. Using this choice of initial conditions, we obtain
\begin{equation}
\begin{split}
\beta_{t,t_0} &\approx \left(\ci\omfreq + \frac{\dot{\omfreq}}{2\ofreq} - \frac{\dot{\omfreq}_0}{2\ofreq_0}\right)e^{-\ci(\bar{\varphi}+\bar{\varphi}_0)}.
\end{split}
\end{equation}

The requirement of a stable vacuum provides a way to define adiabatic initial conditions similar to the originally advocated minimization postulate by~\citet{Parker1969} (see~\cite{Chung2003} for a discussion on the adiabatic vacuum and the intrinsic ambiguity of its definition). Instead of requiring the minimization of the created particle number order by order as in the minimization postulate, we require that the particle number density does not depend on fast variables up to some order. Our requirement can be advantageous when dealing with multiple fields with quadratic actions (only coupled by quadratic terms in the Hamiltonian), because the AA variables can be used in such systems by identifying the fast and slows variables. In principle, this criteria could be used when dealing with non-linear systems since, in this case, it is also possible to define AA variables. 

The above calculation shows that, we can define a stable vacuum order by order by properly choosing the initial conditions. Notwithstanding, the procedure used to obtain the higher order corrections for $\epsilon$ and $\gamma$ is cumbersome and, for higher orders, one must also compute further corrections coming from the angle $\bar{\varphi}$.\footnote{This correction is function of the adiabatic Hannay angle, see for example~\cite{Calkin1996}.} Instead of following this path, in the next sections we use a different approach. We will show that, making a LCT, we can redefine the mass $\om$ and frequency $\ofreq$ of the system, such that  we obtain $\omfreq \propto \OO{\EV{\HFi}^{-n}}$ for these new variables. Then, one can define the leading adiabatic initial condition in the new representation and transform back to the original representation obtaining a stable vacuum of order $\OO{\EV{\HFi}^{-n}}$.

\subsection{Comparison with WKB approach}
\label{sec:cmp}

In the literature, the standard approach for the asymptotic analysis of the solutions is through the WKB approximation. For this, one starts by finding a variable where the field equations have the mass $\om = 1$. Then, the WKB analysis is performed and the solution is translated back to the original variables. Let us consider the simple example of a massless spectator scalar field in a power law inflationary model. For the KG field in Eq.~\eqref{eq:H:KG},\footnote{Here we use the conformal time defined by $\dd{}t = a\dd\eta$ and the prime denotes the derivative with respect to $\eta$.} $\om \equiv a^2$ and $\ofreq^2 \equiv \EV{\HFi}^2$, the field satisfies 
\begin{equation}
\mf^{\prime\prime}_{\HFi} + 2\frac{a^\prime}{a}\mf^\prime_{\HFi} + \EV{\HFi}^2\mf_{\HFi} = 0,
\end{equation}
where changing the field variable to $Q_{\HFi} = a\mf_{\HFi}$, we get 
\begin{equation}
Q^{\prime\prime}_{\HFi} + \left(\EV{\HFi}^2 - \frac{a^{\prime\prime}}{a}\right)Q_{\HFi} = 0.
\end{equation}
In this example we assume that the background evolves with a simple power law, $a(\eta) = (-\eta)^{b+\frac12}$ and a massless field $\mu = 0$.
We can write the analytic solution for the field as
\begin{equation}\label{eq:anal_solut}
Q_{\EV{\HFi}} = \sqrt{-\eta\frac{\pi}{4}}e^{\ci\frac{\pi}{2}\left(b+\frac{1}{2}\right)}H^{(1)}_b(-\EV{\HFi}\eta),
\end{equation}
where $H^{(1)}_b$ is the Hankel function of the first kind. This solution matches the boundary condition 
\begin{equation}
\lim_{\eta\rightarrow-\infty}=\frac{e^{-\ci\EV{\HFi}\eta}}{\sqrt{2\EV{\HFi}}}.
\end{equation}
Naturally, it also matches the AA approximation [see Eq.~\eqref{eq:BT1:leading}] and, since the leading order coincides with the WKB approximation, it also matches the ``positive frequency'' WKB solution.

\begin{figure*}
\includegraphics[scale=0.58]{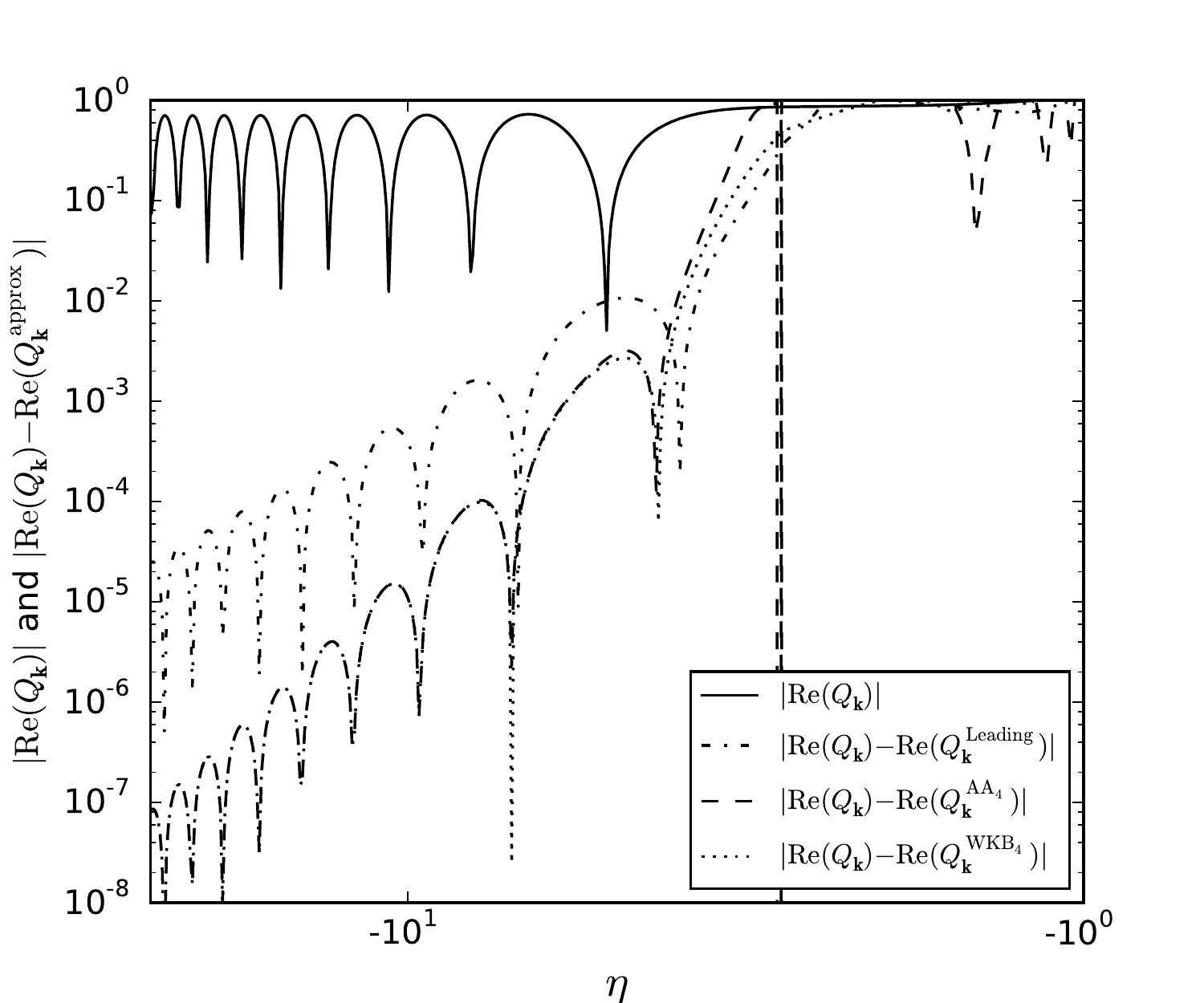}
\includegraphics[scale=0.58]{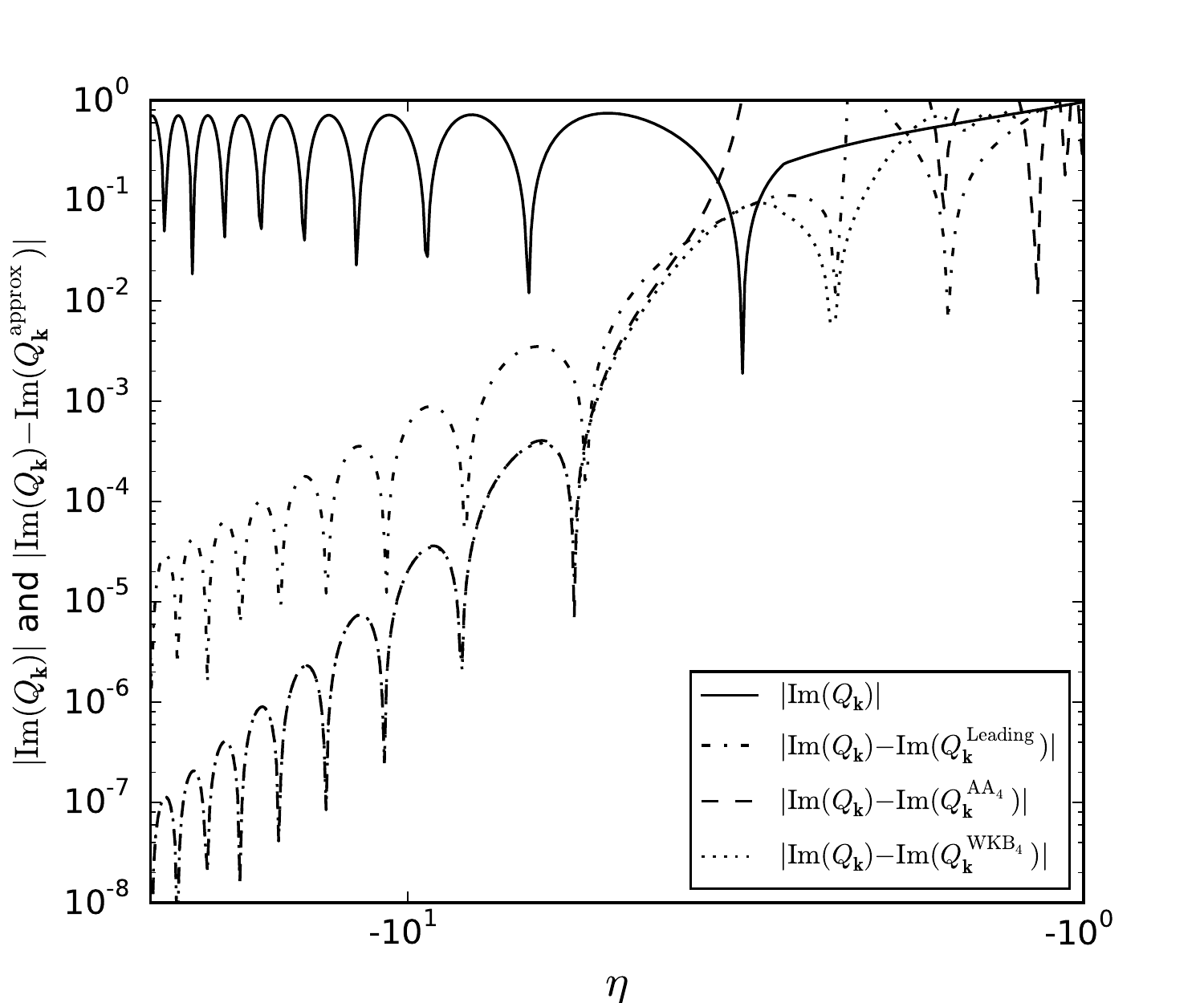}
\includegraphics[scale=0.58]{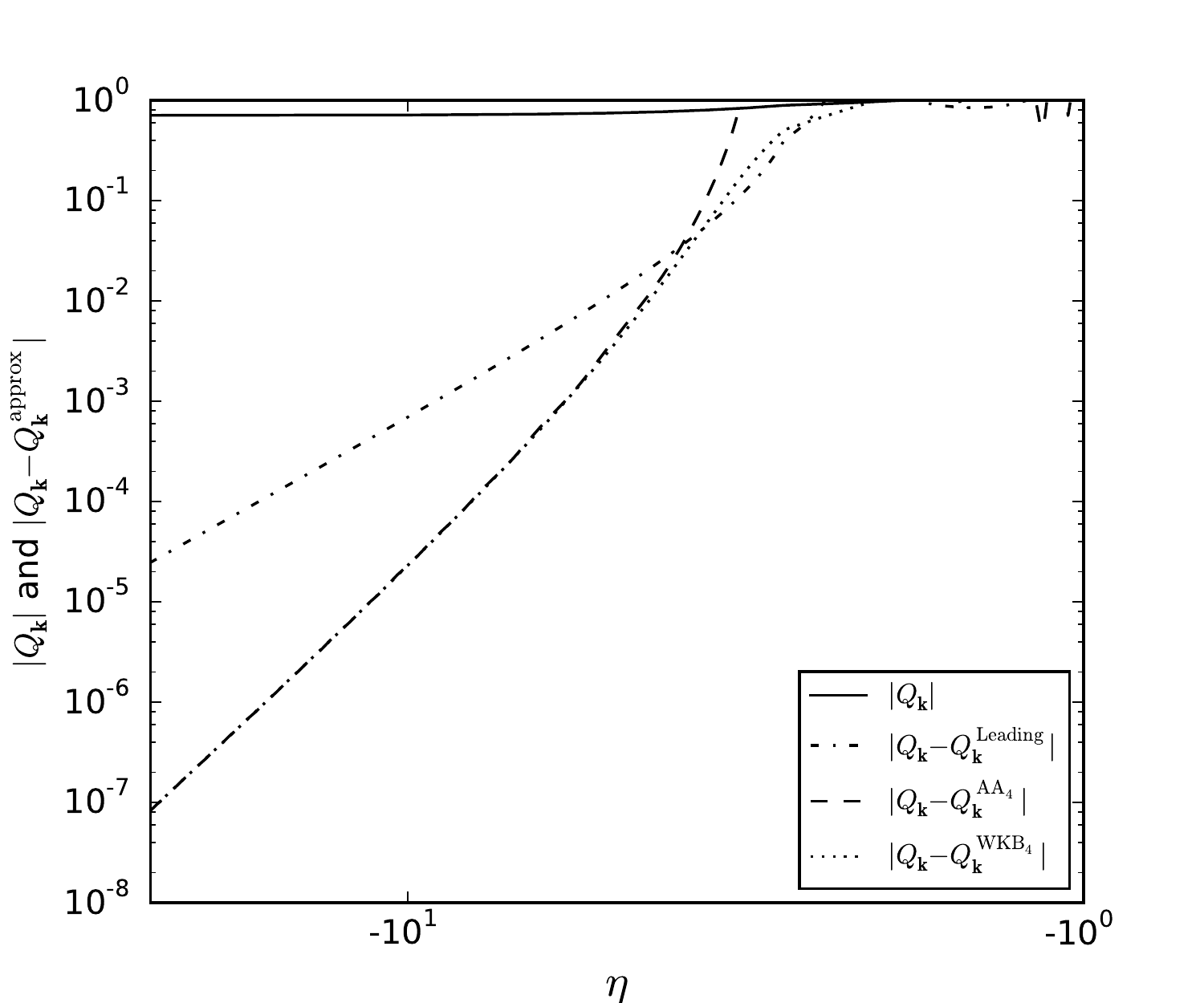}
\includegraphics[scale=0.58]{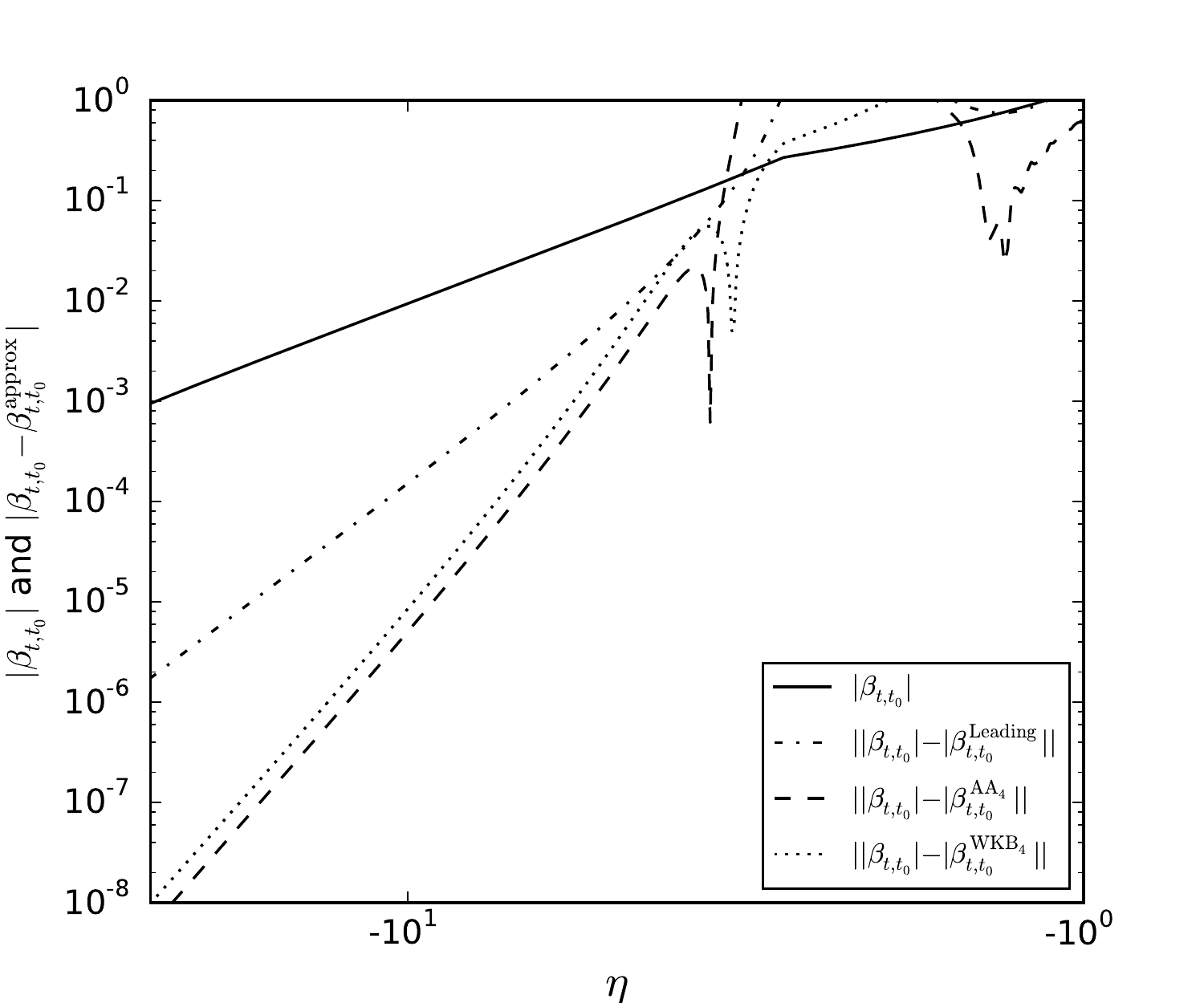}
\caption{\label{fig:basis} Plots of the the basis function $Q_{\EV{\HFi}}$ described in Sec.~\ref{sec:cmp}. The two figures on the top panel represent the real and imaginary parts of the analytic solution and the difference between the analytic solution and the different approximations. The lower left figure shows the absolute value of the analytic function and its difference with the approximations. The lower right describe the time evolution of the $\beta_{t,t_0}$ function for this analytic basis function and again its comparison with the approximations. In this figures we used $\EV{\HFi} = 1$ and $b = -2$, every quantity is written in units of $c/H_0 = ca_0/\dot{a}_0$. The dark vertical lines in the top left plot is the result of spurious oscillations of the AA approximation in the region where the it fails.}
\end{figure*}

In Fig.~\ref{fig:basis} we show the comparison between the analytic solution [Eq.~\eqref{eq:anal_solut}] and three different approximations. The leading order approximation is given by Eq.~\eqref{eq:BT1:leading} and it is equivalent to the WKB approximation. To highlight the difference between the methods, we compute the fourth order WKB and AA corrections. In the two upper panels of Fig.~\ref{fig:basis}, we can note that the approximations present different behaviors when they get closer to the end of their validity range. In the lower left panel, we can see that the analytic solution is proportional to a simple complex exponential,\footnote{This is to be expected by studying the asymptotic series approximation for the Bessel functions.} i.e., its modulus does not oscillate. All three approximations also have this feature, additionally they are also in phase with the analytic solution, i.e., the module of the difference between them and the analytic solution $\left\vert Q_{\EV{\HFi}} - Q_{\EV{\HFi}}^{\text{approx}}\right\vert$ also does not oscillate. 

The main non-trivial difference between WKB and AA is that in the AA approach we obtain approximations for both field and its momentum. In the WKB approach, we can use the approximation for the field and obtain an approximation for the momentum using one of the Hamilton equations, e.g., we could define the momentum as $\Pi_{Q_{\EV{\HFi}}}^{\text{WKB}} \equiv Q_{\EV{\HFi}}^{\text{WKB}\prime}$. However, this implies an additional ambiguity when defining the initial conditions for the Hamilton equations. That is, we could have used any other momentum definition as long as it agrees with the approximation. Using this momentum definition, we also plot the $\beta_{t,t_0}$ function and its comparison with the three approximations. In the right lower panel of Fig.~\ref{fig:basis}, we can note that AA provides a slight better approximation for the $\beta_{t,t_0}$ function than the WKB one.

The WKB approximation is widely used in this context, since it application is more straightforward. In some scenarios however, AA is more advantageous. The AA approach provides an approximation for both field and momentum, which allowed us to study the asymptotic behavior of the $\beta_{t,t_0}$ function in Sec.~\ref{sec:vacuum:stab}. The AA variables can be used even when dealing with multiple free fields coupled only by quadratic terms in the Hamiltonian. In the next section we show how to combine the already well known WKB approximation and the AA variables to obtain a straightforward methodology to approximate the basis functions.

\section{Canonical Transformations}
\label{sec:LCT}

In this work we focus in the linear canonical transformations since they keep the system linear and, therefore, we can still apply the canonical quantization. In this case, the LCT group is that one which keeps the symplectic matrix [Eq.~\eqref{eq:def:SM}] invariant. For a single field, the symplectic group is $\text{Sp}(2,\mathbb{R})$, which is tridimensional with the following generators
\begin{equation}
\mathrm{S}_1 = \left( \begin{array}{cc}
1 & 0 \\
0 & -1 \end{array} \right),\quad \mathrm{S}_2 =  \left( \begin{array}{cc}
0 & 1 \\
0 & 0 \end{array} \right),\quad \mathrm{S}_3 = \left( \begin{array}{cc}
0 & 0 \\
1 & 0 \end{array} \right).
\end{equation}
These three matrices generate transformations, respectively, in the form
\begin{align}\label{eq:canonical:one}
\mf_{\HFi} \rightarrow e^{g_1}\mf_{\HFi}, \qquad &\pmf{}_{\HFi} \rightarrow e^{-g_1}\pmf{}_{\HFi}, \\ \label{eq:canonical:two}
\mf_{\HFi} \rightarrow \mf_{\HFi} + g_3 \pmf{}_{\HFi}, \qquad &\pmf{}_{\HFi} \rightarrow \pmf{}_{\HFi}, \\ \label{eq:canonical:three}
\mf_{\HFi} \rightarrow \mf_{\HFi}, \qquad &\pmf{}_{\HFi} \rightarrow \pmf{}_{\HFi} + g_2 \mf{}_{\HFi},
\end{align}
where $g_1$, $g_2$ and $g_3$ are the group parameters related to $\mathrm{S}_1$, $\mathrm{S}_2$ and $\mathrm{S}_3$, respectively.

The first important result is that the annihilation and creation operators, as in Eq.~\eqref{eq:def:AO:fs}, are invariant under LCT. It is easy to see that every product in the form $\BRi_{a}\BSi^{a}$ is scalar under canonical transformations if taken in the same spatial section, i.e., at the same time slice. However, this also means that the annihilation operator in the Heisenberg [Eq.~\eqref{eq:def:AO:heisenberg}] representation is not invariant under canonical transformations, with the exception of the time-independent canonical transformations.

\subsection{Existence}
\label{sec:exists}

In Sec.~\ref{sec:asymp:beta} we showed that, for a Hamiltonian with no cross term, we need to impose conditions on $\omfreq$ to obtain representations with unitary evolution. We show now that, using the original variables, it is not possible to find a representation where the evolution is unitarily implemented.
We have to return to the original canonical variables inverting the transformation in Eq.~\eqref{eq:lct:rm:h}, i.e., 
\begin{equation}
p \rightarrow P = p - \om{}h q, \qquad q \rightarrow Q = q.
\end{equation}
In this canonical representation the phase vector is 
\begin{equation}
v_a \doteq (\BT_1, \BT_2 - \om{}h\BT_1).
\end{equation}
Consequently, the $\beta_{t,t_0}$ function is also modified, and this modification is straightforward to calculate, namely,
\begin{align}
\beta^-_{t,t_0} &\rightarrow \beta^-_{t,t_0} - \frac{\Delta{}h}{2}F^c\left(\ci{}e^{-\ci\bar{\varphi}_0}F^c_0+e^{\ci\bar{\varphi}_0}F^s_0\right), \\
\beta^+_{t,t_0} &\rightarrow \beta^+_{t,t_0} - \frac{\Delta{}h}{2}F^s\left(e^{-\ci\bar{\varphi}_0}F^c_0-\ci{}e^{\ci\bar{\varphi}_0}F^s_0\right),
\end{align}
where we defined 
\begin{equation}
\Delta{}h \equiv \frac{(\om{}h-\om_0h_0)}{2\sqrt{\om_0\ofreq_0\om\ofreq}}.
\end{equation}
The modified matrix, equivalent to Eq.~\eqref{eq:det:beta} but in the new representation, is
\begin{equation}
\left(\begin{array}{cc}
F^c\left[\sinh(\omfreq)-\ci\Delta{}h\right] &  F^c\left[-\ci\cosh(\omfreq) - \Delta{}h\right] \\
F^s\left[\ci\cosh(\omfreq) - \Delta{}h\right] & F^s\left[\sinh(\omfreq) + \ci\Delta{}h\right] \end{array}\right).
\end{equation}
It is easy to check that its determinant is independent of $\Delta{}h$ and, therefore, the same conclusions as those discussed below Eqs.~\eqref{eq:cond:e0g0} follow in this new representation. The limits given in Eqs.~\eqref{eq:lim:beta:pm} are also unmodified, since 
\begin{equation}
\lim_{\EV{\HFi}\rightarrow\infty}\Delta{}h = 0.
\end{equation}
The quantity $\Delta{}h$ must have this limit because we have already required that the curves defined by Eq.~\eqref{eq:E:curves} were closed. In other words, this imposes that $h^2$ should go faster to zero than $\ofreq^2$ and, consequently,
\begin{equation}
\lim_{\EV{\HFi}\rightarrow\infty}\Delta{}h \propto \EV{\HFi}^{-1+n_h},
\end{equation}
where $n_h$ is the largest power of $\EV{\HFi}$ in $h$; it satisfies $n_h < 1$. 

The last requirement is $\lim_{\EV{\HFi}\rightarrow \infty}\EV{\HFi}^3\vert\beta_{t,t_0}\vert^2 = 0$, so we have an additional condition for this representation,
\begin{equation}
\lim_{\EV{\HFi}\rightarrow \infty}\EV{\HFi}^{\frac{3}{2}}\Delta{}h \propto \lim_{\EV{\HFi}\rightarrow \infty}\EV{\HFi}^{\frac{1}{2} + n_h} = 0.
\end{equation}
If $h$ does not contain any spatial derivative operator,\footnote{If $h$ contains any spatial derivative, it must be such that the spectral dependency on the eigenvalue is $\EV{\HFi}^{n_h}$, where it must satisfy $n_h < -1/2$. This would require the appearance of unusual powers of inverse Laplacian in the cross term $qp$ of the Hamiltonian. For simplicity we ignore these cases here.} then $n_h = 0$ and the above condition is impossible to satisfy. This means that the time evolution for the field operators is not unitarily implemented in this new representation. However, a simple LCT, removing the term $hqp$ from the Hamiltonian, is sufficient to change the canonical representation to one where the evolution is unitarily implemented.

We can conclude from these results that it is possible to obtain a canonical representation where the evolution of the field operators is unitarily implemented for an initial Hamiltonian, where the only conditions on the functions $\om$, $\ofreq$ and $h$ are the following: 
\begin{itemize}
\item The frequency $\ofreq$ must satisfy
\begin{equation}\label{eq:cond1}
\lim_{\EV{\HFi}\rightarrow \infty} \ofreq \propto +\EV{\HFi},
\end{equation}
\item The mass $\om$ must be such that the variable $\omfreq$ defined in Eq.~\eqref{eq:def:zeta} satisfies 
\begin{equation}\label{eq:cond2}
\lim_{\EV{\HFi}\rightarrow \infty}\EV{\HFi}^{\frac{3}{2}}\omfreq = 0,
\end{equation}
\item The cross term $h$ must satisfy 
\begin{equation}\label{eq:cond3}
\lim_{\EV{\HFi}\rightarrow \infty}(\ofreq^2 - h^2) > 0.
\end{equation}
\end{itemize}
This canonical representation can be obtained explicitly using the transformations defined by Eqs.~\eqref{eq:lct:rm:h}, \eqref{eq:redef:freq}.

\subsection{Uniqueness}
\label{sec:unique}

We have already shown that a canonical representation with unitary evolution exists. Now we can show that this representation is unique. Consider a general canonical transformation parametrized as 
\begin{equation}\label{eq:def:gen:LCT}
C_a{}^b \doteq e^{g_1\mathrm{S}_1}e^{g_2\mathrm{S}_2}e^{g_3\mathrm{S}_3} = \left[
\begin{array}{cc}
 e^{g_1} (g_2g_3 + 1) & e^{g_1} g_2 \\
 e^{-g_1} g_3 & e^{-g_1} \\
\end{array}
\right],
\end{equation}
where $g_1$, $g_2$ and $g_3$ are functions of the time. At this point we consider only global transformations, i.e., we assume these functions do not depend on the eigenvalue $\EV{\HFi}$. We also require that the transformations are real, otherwise the new field operators would fail to be Hermitian. These transformations do not alter the meaning of the field, in the sense that they merely multiply the field by the same value at every point in the same time slice.

It is easy to check that $C_a{}^c(t)\SM^{ab}C_b{}^d(t) = \SM^{cd}$, i.e., two transformations at the same time slice maintain $\SM^{ab}$ invariant. However, as we noted before, the Heisenberg creation and annihilation operators are not invariant under such transformations, as they are defined by products of canonical variables at different time slices. We can check the consequences of this transformation at the time evolution by inspecting the transformed $\beta^C_{t,t_0}$ function [Eq.~\eqref{eq:alpha:beta:t}] in this new representation, i.e.,
\begin{equation}\label{eq:beta:trans}
\beta^C_{t,t_0} = \BTi_c C^{cd}_{t_0,t}\BT_d, \quad C_{t_0,t}^{cd} \equiv C_a{}^c(t_0)\SM^{ab}C_b{}^d(t),
\end{equation}
or more explicitly
\begin{equation}
\beta^C_{t,t_0} = \BTi_1\BT_1 C^{11}_{t_0,t} + \BTi_2\BT_2 C^{22}_{t_0,t} + \BTi_1\BT_2 C^{12}_{t_0,t} + \BTi_2\BT_1 C^{21}_{t_0,t}.
\end{equation}
Starting from an unitarily implemented representation, the expressions for $\BT_a$ and $\BTi_a$ are given by Eqs.~\eqref{eq:BT1} and \eqref{eq:BT2}. From these equations we assert that 
\begin{align}
\lim_{\EV{\HFi}\rightarrow \infty}\BT_1 &\propto \lim_{\EV{\HFi}\rightarrow \infty}\BTi_1 \propto \EV{\HFi}^{-\frac{1}{2}}, \\
\lim_{\EV{\HFi}\rightarrow \infty}\BT_2 &\propto \lim_{\EV{\HFi}\rightarrow \infty}\BTi_2 \propto \EV{\HFi}^{\frac{1}{2}}.
\end{align}
Consequently, we have tree different spectral dependencies in the function $\beta^C_{t,t_0}$. In order to satisfy $\lim_{\EV{\HFi}\rightarrow \infty}\EV{\HFi}^3\vert\beta^C_{t,t_0}\vert^2 = 0$, we must impose that 
\begin{equation}
C^{11}_{t_0,t} = 0 = C^{22}_{t_0,t}.
\end{equation}
As we are assuming that the original representation has a unitary evolution, then 
\begin{equation}
\lim_{\EV{\HFi}\rightarrow \infty}\vert\beta_{t,t_0}\vert^2 = \ci(\BTi_1\BT_2-\BTi_2\BT_1) = 0,
\end{equation}
which results in 
\begin{equation}
\lim_{\EV{\HFi}\rightarrow \infty}\BTi_1\BT_2 = \BTi_2\BT_1.
\end{equation}
Hence we must also impose that 
\begin{equation}
C^{12}_{t_0,t} + C^{21}_{t_0,t} = 0.
\end{equation}
Writing $C^{ab}_{t_0,t}$ explicitly, we get
\begin{align}\nonumber
C^{11}_{t_0,t} &= \ci{}g_3(t) e^{g_1(t_0) - g_1(t)} \left[g_2 (t_0) g_3 (t_0) + 1\right] \\
&- \ci{}g_3 (t_0) \left[g_2 (t) g_3 (t) + 1\right] e^{g_1 (t) - g_1 (t_0)}, \\ 
C^{22}_{t_0,t} &= \ci{}g_2 (t_0) e^{g_1 (t_0)- g_1 (t)} - \ci g_2 (t) e^{g_1 (t) - g_1 (t_0)}, \\ \nonumber
C^{12}_{t_0,t} &= \ci{}e^{g_1 (t_0)- g_1 (t)} \left[g_2 (t_0) g_3 (t_0) + 1\right] \\ 
&- \ci{}g_2 (t) g_3 (t_0) e^{g_1 (t) - g_1 (t_0)}, \\ \nonumber
C^{21}_{t_0,t} &= \ci{}g_3 (t) g_2 (t_0) e^{g_1 (t_0)-g_1 (t)}\\ 
&-\ci{}\left[g_2 (t) g_3 (t) + 1\right] e^{g_1 (t)-g_1 (t_0)}.
\end{align}
We can first solve $C^{11}_{t_0,t} = 0 = C^{22}_{t_0,t}$ to obtain $g_2(t)$ and $g_3(t)$ in terms of $g_1(t)$, i.e.,
\begin{align}
g_2(t) &= g_2 (t_0) e^{-2 [g_1 (t)-g_1 (t_0)]}, \\ 
g_3(t) &= g_3 (t_0) e^{2 [g_1 (t)-g_1 (t_0)]}.
\end{align}
Substituting them back into the final condition $C^{12}_{t_0,t} + C^{21}_{t_0,t} = 0$, we obtain
\begin{equation}
\sinh [g_1 (t)-g_1 (t_0)] = 0.
\end{equation}
The only solution is $g_1(t)$ constant, which consequently implies that $g_2(t)$ and $g_3(t)$ are also constant. Therefore, there is no time dependent LCT connecting to another set of canonical variables where the time evolution can be represented by unitary operators. In other words, canonical variables found in Sec.~\ref{sec:exists} are unique.

\subsection{The background conditions}
\label{sec:back:cond}

In the Sec.~\ref{sec:asymp:beta} we had to assume that $\omfreq$ satisfies the condition given in Eq.~\eqref{eq:omfreq:lim} to obtain an unitary evolution for the field operators. From the existence conditions, we realized that only the representations where the Hamiltonian does not contain a cross term implement an unitary evolution. The canonical transformation, applied to remove the cross term [Eq.~\eqref{eq:lct:rm:h}], modifies the Hamiltonian through 
\begin{equation}
\om \rightarrow \om, \quad h\rightarrow 0, \quad \ofreq^2 \rightarrow \ofreq^2 - h^2 - \frac{\lie_\n(h\om)}{\om}.
\end{equation}
Hence, assuming that initially $\omfreq$ does not satisfy the limit in Eq.~\eqref{eq:omfreq:lim}, we can search for canonical transformations, such that the new product $\om\ofreq$ fulfills the cited condition. Starting from a representation with $h = 0$, we must obtain the subset of canonical transformations which preserves $h = 0$.

Before studying these transformations, it is informative to look if a different choice of the time coordinate could be used to transform $\omfreq$ to satisfy the unitary condition. To evaluate this, we first note that it is possible to change the time variable such that $\dd{}t = \lapse\dd\tau$, where $t$ is the current time variable, $\tau$ the new one and $N$ an arbitrary lapse function. Examining the action
\begin{equation}
\begin{split}
S = \frac{1}{2}\int\dd^4x\left(\ci\chi_a\SM^{ab}\dot{\chi}_b - \chi_a\gH^{ab}\chi_b\right),
\end{split}
\end{equation}
one notes that the first term will simply change as 
\begin{equation}
\dd{}t\dot{\chi}_b \rightarrow \dd\tau\chi^\prime_b,
\end{equation}
where new a prime represents a derivative with respect to $\tau$, and the second term gains a factor of $\lapse$, i.e., $\gH^{ab} \rightarrow \lapse\gH^{ab}$. Finally, mass and frequency transform as 
\begin{equation}\label{eq:time:gauge}
\om \rightarrow \frac{\om}{\lapse},\qquad \ofreq \rightarrow \lapse \ofreq,
\end{equation}
and, consequently, $\omfreq$ is invariant under the change of time coordinate. Therefore, the appropriate canonical variable is independent of the time gauge choice.

The general LCT expressed in Eq.~\eqref{eq:def:gen:LCT} maps the canonical variables $\BT_a$ into $\BR_a$ as 
\begin{equation}
\BR_a = C_a{}^b\BT_b, \qquad \BT_a = \left(C_a{}^b\right)^{-1}\BR_b,
\end{equation}
where
\begin{equation}
\begin{split}
\left(C_a{}^b\right)^{-1} &\doteq e^{-g_3\mathrm{S}_3}e^{-g_2\mathrm{S}_2}e^{-g_1\mathrm{S}_1} \\
&= \left[
\begin{array}{cc}
 e^{-g_1} & -e^{g_1} g_2 \\
 -e^{-g_1} g_3 & e^{g_1} (g_2g_3 + 1) \\
\end{array}
\right].
\end{split}
\end{equation}
The calculation can be greatly simplified using the parameterization 
\begin{equation}
g_1 = \ln(\sqrt{\om/\onm{1}}), \quad g_2 = 0,
\end{equation}
where $\onm{1}$ is an arbitrary positive function.\footnote{A LCT with $g_2 \neq 0$ would result in a cross term proportional to $\ofreq^2{}g_2$. Therefore, to remove such terms while keeping $g_2 \neq 0$, one must choose $g_1$, $g_2$ or $g_3$ as functions of $\om$ and $\ofreq$. This however makes the transformation dependent on $\EV{\HFi}$, which is exactly what we want to avoid.} This transformation, when applied to the Hamiltonian, results in
\begin{align}
\gH(\BR) &= \frac{1}{2}\BR_a\gH_{c_{1}}^{ab}\BR_b,\\ 
\gH_{c_{1}}^{ab} &\equiv \left[ \begin{array}{cc}
\onm{1}\left(\ofreq^2 + {\frac{g_3^2}{\om^2}} - \frac{\dot{g}_3}{\om}\right) & \lie_\n\ln\left(\sqrt{\frac{\om}{\onm{1}}}\right) - \frac{g_3}{\om} \\
\lie_\n\ln\left(\sqrt{\frac{\om}{\onm{1}}}\right) - \frac{g_3}{\om} & \frac{1}{\onm{1}} \end{array} \right].
\end{align}
Using the function $g_3$ to remove the cross terms, we finally obtain
\begin{equation}
\gH_{c_{1}}^{ab} \equiv \left[ \begin{array}{cc}
\onm{1}\ofreq^2 - \frac{1}{\sqrt{\frac{\om}{\onm{1}}}}\lie_\n\left(\onm{1}\dot{\sqrt{\frac{\om}{\onm{1}}}}\right) & 0 \\
0 & \frac{1}{\onm{1}} \end{array} \right].
\end{equation}
Thus, under this canonical transformation, the mass and frequency of the Hamiltonian are transformed as
\begin{align}
\om &\rightarrow \onm{1}, \\
\ofreq^2 &\rightarrow \onfreq{1}^2 \equiv \ofreq^2 - \frac{1}{\onm{1}\sqrt{\frac{\om}{\onm{1}}}}\lie_\n\left(\onm{1}\dot{\sqrt{\frac{\om}{\onm{1}}}}\right),
\end{align}
where $\onm{1}$ represents the only degree of freedom left in the LCT and is an arbitrary function of time only. Note also that the transformed frequency can be written as
\begin{equation}\label{eq:c1:c2}
\onfreq{1}^2 = \ofreq^2 - \frac{\ddot{\sqrt{\om}}}{\sqrt{\om}} + \frac{\ddot{\sqrt{\onm{1}}}}{\sqrt{\onm{1}}}.
\end{equation}

After performing the LCT, one can make the same asymptotic analysis described in Sec.~\ref{sec:asymp:beta}. But now the product given in Eq.~\eqref{eq:def:mv} is 
\begin{equation}
(\onm{1}\onfreq{1})^2 = \sum_{i=-2}^{\infty}f_{c_1,i}(t)\EV{\HFi}^{-i},
\end{equation}
where $$f_{c_1,i}(t) \equiv \frac{\onm{1}^2}{\om^2} f_{i}(t) - \delta_{i,0}\frac{\onm{1}}{\sqrt{\frac{\om}{\onm{1}}}}\lie_\n\left(\onm{1}\dot{\sqrt{\frac{\om}{\onm{1}}}}\right).$$ For most field theories, including those where the field has a canonical kinetic term, the only non-zero $f_i(t)$ are $f_{-2}(t)$ and $f_{0}(t)$. So in these cases we can choose 
\begin{equation}\label{eq:mfreq:cond}
\onm{1} = \frac{\om}{\sqrt{f_{-2}(t)}},
\end{equation}
making the new 
\begin{equation}
\onmfreq{1} \equiv \ln\left\{\sqrt{\onm{1}\onfreq{1}/[\onm{1}(t_0)\onfreq{1}(t_0)]}\right\}
\end{equation}
to fulfill the unitary evolution necessary conditions. In a general case, this transformation is sufficient if $f_{-1}(t) \propto f_{-2}(t)$ or $f_{-1}(t) = 0$. 

In short, the LCT generated by 
\begin{equation}\label{eq:u:lct}
\begin{split}
g_1 &= \ln\left(\sqrt{\frac{\om}{\onm{1}}}\right), \quad g_2 = 0, \\ 
g_3 &= \om\lie_\n\left[\ln\left(\sqrt{\frac{\om}{\onm{1}}}\right)\right], \quad \onm{1} = \frac{\om}{\sqrt{f_{-2}(t)}},
\end{split}
\end{equation}
leads to a choice of canonical variables for the fields, where it is possible to implement the time evolution of the quantum fields unitarily. We denote this canonical representation as $c_1$ and the basis defined in these canonical variables as $\BR_a^{c_1}$.

The matrix describing the above transformation is 
\begin{equation}
\begin{split}
C_a{}^b &\doteq \left[
\begin{array}{cc}
\sqrt{\frac{\om}{\onm{1}}} & 0 \\
\onm{1}\lie_\n\left(\sqrt{\frac{\om}{\onm{1}}}\right) & \sqrt{\frac{\onm{1}}{\om}} \\
\end{array}
\right].
\end{split}
\end{equation}
Applying it to the field variables, we obtain 
\begin{equation}
\mf_{\HFi} \rightarrow \sqrt{\frac{\om}{\onm{1}}}\mf_{\HFi}, \quad \pmf{}_{\HFi} \rightarrow \sqrt{\frac{\onm{1}}{\om}}\pmf{}_{\HFi} + \onm{1}\lie_\n\sqrt{\frac{\om}{\onm{1}}}\mf{}_{\HFi}.
\end{equation}
Therefore, the field variable is changed by a simple time-dependent re-scaling. In the new canonical representation the field basis is given by Eq.~\eqref{eq:BT1}, trading the mass and frequency for the new ones. In the UV limit the basis functions are 
\begin{equation}
\lim_{\EV{\HFi}\rightarrow \infty}\BR_1^{c_1} \approx \frac{1}{2}\sqrt{\frac{1}{\onm{1}\onfreq{1}}}e^{-\ci\bar{\varphi}}(1+\ci) = \frac{e^{-\ci\bar{\varphi}+\ci\pi/4}}{\sqrt{2\EV{\HFi}}},
\end{equation}
where we assumed that the initial conditions were such that the vacuum is stable at least at leading order [Eq.~\eqref{eq:uv:init}]. The same basis written in the original variable is
\begin{equation}
\lim_{\EV{\HFi}\rightarrow \infty}\BT_1 \approx \frac{e^{-\ci\bar{\varphi}+\ci\pi/4}}{\sqrt{2\sqrt{f_{-2}(t)}\EV{\HFi}}}.
\end{equation}
In the representation $c_1$, the basis functions $\BR_a^{c_1}$ have a very simple evolution, in the UV limit,  described by the time evolution of the phase. This is exactly the evolution of the basis functions of a free field in the Minkowsky space-time. On the other hand, in the original representation, the basis functions have a more complicated evolution emanating from the $f_{-2}(t)$ term. 

In light of this discussion, it is worth reevaluating the example in Sec.~\ref{sec:cmp}. In the original variables, we saw that the mass and frequency are $\om = a^2$ and $\ofreq = \EV{\HFi}$, respectively, giving  $$\om^2\ofreq^2 = a^4\EV{\HFi}^2,$$ where we identify $f_{-2} = a^4$ and $f_{i} = 0$ for all $i \neq -2$. Applying the condition in Eq.~\eqref{eq:mfreq:cond} we obtain the new mass $\onm{1} = 1$. This shows that the $Q_{\EV{\HFi}}$ variable is actually the right variable for quantization with unitary evolution. 

In cosmology it is a common practice to parametrize the field such that its action resembles that of a free field in Minkowsky space-time with a time-dependent mass term (see for example~\cite{Mukhanov2005}). In the language presented in this work, this means changing the mass variable to a new $\onm{1} = 1$ and choosing the conformal time gauge. The fact that the new mass is constant is, however, just a consequence of the time gauge choice. In the initial time gauge, the mass and frequency are, respectively, $\om = a^3$ and $\ofreq = a^{-1}\EV{\HFi}$. Within this choice we obtain $f_{-2} = a^4$, which is expected since we showed that this product is invariant under time gauge change. On the other hand, the new mass is $\onm{1} = a^{-1}$, again compatible with the transformation in Eq.~\eqref{eq:time:gauge}. 

Consequently, there is nothing special about casting the field as a field with time-dependent mass in a Minkowsky background. This is possible when the new mass $\onm{1}$ is simple enough to be made constant by a time gauge choice. For a free field in a FLRW background, the Laplacian operator scales as $a^{-2}$ and, thus, the frequency contains a term of the form $\ofreq = a^{-1}\EV{\HFi}$. So, a simple choice of the lapse function $N = a$ modifies the frequency such that its part containing the Laplacian is now constant. In this case, it is clear by Eq.~\eqref{eq:mfreq:cond} that the $c_1$ mass is also constant. Nevertheless, this is not always the case, in different settings in cosmology,\footnote{For example, K-inflation~\cite{Garriga1999} or fluid quantization~\cite{Vitenti2013}.} the  Laplacian operator comes with the speed of sound of the perturbed matter content. For these latter examples, the time gauge will also depend on the speed of sound, if it is non-constant. Also, if there are several free fields with different speeds of sound, there is not a single time gauge choice where all $c_1$ masses are constant. 

In physical terms, this means that in the original canonical variables, the UV modes feel the influence of the background evolution at all scales. If one considers FLRW as the background metric, which has homogeneous and isotropic hypersurfaces, it should be expected that the modes that are deep in the UV limit should not be influenced by the background evolution. Hence, the LCT above selects the only canonical variables for the field for which it is possible to find a representation for the quantum fields such that the basis functions in the UV limit evolve as those of a free field in the Minkowsky space-time. Besides, this choice of canonical variables is the only one where the time evolution of the quantum field can be implemented unitarily. We argue that, the equivalence principle is satisfied if the basis functions, in the UV limit, reduce to those of a free field in Minkowsky space-time. For any other choice of canonical variables, it is impossible to find a representation such that its basis satisfies the equivalence principle. In this sense, one can summarize this result in the following statement: The only variable for which we can apply the equivalence principle is also the only variable for which we can represent the field such that its evolution is unitary.

\subsection{Higher order approximations}
\label{sec:higher}

In Sec.~\ref{sec:vacuum:stab} we showed how to obtain a leading order and a $\OO{\EV{\HFi}^{-3}}$ stable vacuum. The calculation necessary to determine such vacuum is involved and requires a careful asymptotic expansion of different integrals. Note that the first correction for the leading order stable vacuum comes from the choice of initial conditions in Eq.~\eqref{eq:first:stable}. Hence, the correction is proportional to the term $\dot{\omfreq}/\ofreq$. 

Instead of computing the adiabatic expansion explicitly, we propose an alternative and straightforward method to obtain higher order approximations and higher order stable vacuum definitions. Starting from the canonical representation satisfying the UV condition, we know from Eq.~\eqref{eq:omfreq:lim} that $\donmfreq{1} \propto \OO{\EV{\HFi}^{-2}}$. Therefore, if we find a new representation such that the new $\onmfreq{2}$ drops faster than $\onmfreq{1}$, then the leading adiabatic approximation in this new representation will have an error of the order $\donmfreq{2}/\onfreq{2}$, instead of the original $\donmfreq{1}/\onfreq{1} \propto \OO{\EV{\HFi}^{-3}}$.

Performing a new transformation in the form of Eq.~\eqref{eq:u:lct} we obtain the new mass and frequency
\begin{align}
\onm{1} &\rightarrow \onm{2}, \\
\onfreq{1}^2 &\rightarrow \onfreq{2}^2 \equiv \onfreq{1}^2 - \frac{1}{\onm{2}\sqrt{\frac{\onm{1}}{\onm{2}}}}\lie_\n\left(\onm{2}\dot{\sqrt{\frac{\onm{1}}{\onm{2}}}}\right).
\end{align}
The adiabatic approximation is controlled by the factor $\onm{2}\onfreq{2}$ appearing in the function 
\begin{equation}
\onmfreq{2} \equiv \ln\left(\sqrt{\frac{\onm{2}\onfreq{2}}{\onm{2}{}_0\onfreq{2}{}_0}}\right),
\end{equation}
where the new mass $\onm{2}$ is an arbitrary function. If we choose $\onm{2} = \EV{\HFi}/\onfreq{1}$, we get
\begin{equation}\label{eq:m_vu_c2}
\begin{split}
\onfreq{2}^2 &= \onfreq{1}^2\left[1 - \frac{1}{\onfreq{1}}\lie_\n\left(\frac{\donmfreq{1}}{\onfreq{1}}\right) - \frac{\donmfreq{1}^2}{\onfreq{1}^2}\right], \\
\onm{2}^2\onfreq{2}^2 &= \EV{\HFi}\left[1 - \frac{1}{\onfreq{1}}\lie_\n\left(\frac{\donmfreq{1}}{\onfreq{1}}\right) - \frac{\donmfreq{1}^2}{\onfreq{1}^2}\right].
\end{split}
\end{equation}
This new canonical representation is labeled $c_2$. Since $\donmfreq{1}/\onfreq{1} \propto \OO{\EV{\HFi}^{-3}}$, we obtain from Eq.~\eqref{eq:m_vu_c2} 
\begin{equation}
\begin{split}
&\lim_{\EV{\HFi}\rightarrow \infty}\onfreq{2} \propto \OO{\EV{\HFi}^{1}}, \\
&\lim_{\EV{\HFi}\rightarrow \infty}\onmfreq{2} \propto \donmfreq{2} \propto \OO{\EV{\HFi}^{-4}}.
\end{split}
\end{equation}
Now, using the leading order adiabatic approximation [equivalent to Eqs.~\eqref{eq:BT1:leading} and \eqref{eq:BT2:leading}], we have
\begin{align}
\BR_1^{c_2} &= \frac{e^{-\ci\int\limits_{t_0}^t\onfreq{2}{}(t_1)\dd{}t_1}}{\sqrt{2\onm{2}\onfreq{2}}} + \OO{\frac{\donmfreq{2}}{\onfreq{2}}}, \\ 
\BR_2^{c_2} &= -\ci\sqrt{\frac{\onm{2}\onfreq{2}}{2}}e^{-\ci\int\limits_{t_0}^t\onfreq{2}(t_1)\dd{}t_1} + \OO{\frac{\donmfreq{2}}{\onfreq{2}}}.
\end{align}
Hence, using the above approximations, the error is of order $\OO{\EV{\HFi}^{-5}}$. Since the LCT used to get the $c_2$ representation is exact, we can transform back to the original variables obtaining an approximation with an error of order $\OO{\EV{\HFi}^{-5}}$, namely,
\begin{align}
\BR_1^{c_1} &= \frac{e^{-\ci\int\limits_{t_0}^t\onfreq{2}{}(t_1)\dd{}t_1}}{\sqrt{2\onm{1}\onfreq{2}}} + \OO{\frac{\donmfreq{2}}{\onfreq{2}}}, \\
\BR_2^{c_1} &= -\sqrt{\frac{\onm{1}\onfreq{2}}{2}}\left(\ci+\frac{\donmfreq{1}}{\onfreq{2}}\right)e^{-\ci\int\limits_{t_0}^t\onfreq{2}(t_1)\dd{}t_1} + \OO{\frac{\donmfreq{2}}{\onfreq{2}}}.
\end{align}
Note that these equations are not the leading order adiabatic approximation using the variables $\BR^{c_1}_a$, but the leading order approximation obtained using the variables $\BR^{c_2}_a$ transformed back to the $\BR^{c_1}_a$ variables. The transformation matrix between these two sets of variables is
\begin{equation}
\begin{split}
C_a{}^b(c_1\rightarrow c_2) &\doteq \left(
\begin{array}{cc}
\sqrt{\frac{\onm{1}\onfreq{1}}{\EV{\HFi}}} & 0 \\
\sqrt{\EV{\HFi}\onm{1}\onfreq{1}}\frac{\donmfreq{1}}{\onfreq{1}} & \sqrt{\frac{\EV{\HFi}}{\onm{1}\onfreq{1}}} \\
\end{array}
\right).
\end{split}
\end{equation}
We already know that the leading order solution in the $c_2$ representation provides a unitary evolution for the quantum fields in the same $c_2$ representation. Now we must show that the leading solution in the $c_2$ representation, when transformed back to the $c_1$ representation, also provides a unitary evolution for the fields in the $c_1$ representation. To evaluate this aspect, we must calculate the number of particles measured between these two representations [see Eq.~\eqref{eq:def:alpha:beta}],
\begin{equation}
\beta_{c_2,c_1} \equiv \BR_a^{c_1}\SM^{ab}\BR_b^{c_2}.
\end{equation}
Using the adiabatic approximations, we obtain
\begin{equation}
\begin{split}
\lim_{\EV{\HFi}\rightarrow \infty}\beta_{c_2,c_1} &\approx \frac{1}{2}\left(\sqrt{\frac{\EV{\HFi}}{\onm{1}\onfreq{1}}}-\sqrt{\frac{\onm{1}\onfreq{1}}{\EV{\HFi}}}\right)\\
&+\frac{\ci}{2}\sqrt{\frac{\onm{1}}{\EV{\HFi}\onfreq{2}}}\donmfreq{1} = \OO{\EV{\HFi}^{-2}}.
\end{split}
\end{equation}
Consequently, the two representations $c_1$ and $c_2$ are unitarily equivalent. 

Repeating the same procedure, but using 
\begin{equation}
\onm{n} = \frac{\EV{|HFi}}{\onfreq{n-1}},
\end{equation}
we obtain the general result
\begin{equation}
\begin{split}
&\lim_{\EV{\HFi}\rightarrow \infty}\onfreq{n} \propto \OO{\EV{\HFi}^{1}}, \\
&\lim_{\EV{\HFi}\rightarrow \infty}\onmfreq{n} \propto \donmfreq{n} \propto \OO{\EV{\HFi}^{-2n}}.
\end{split}
\end{equation}
The canonical transformation matrix connecting these representations with $c_1$ is
\begin{equation}
\begin{split}
C_a{}^b(c_1\rightarrow c_{n}) &\doteq \left(
\begin{array}{cc}
\sqrt{\frac{\onm{1}}{\onm{n}}} & 0 \\
\sqrt{\onm{1}\onm{n}}\sum_{l=1}^{n-1}\donmfreq{l} & \sqrt{\frac{\onm{n}}{\onm{1}}} \\
\end{array}
\right), \\
&\doteq \left(
\begin{array}{cc}
\sqrt{\frac{\onm{1}\onfreq{n-1}}{\EV{\HFi}}} & 0 \\
\frac{\lie_\n{\sqrt{\EV{\HFi}\onm{1}\onfreq{n-1}}}}{\onfreq{n-1}} & \sqrt{\frac{\EV{\HFi}}{\onm{1}\onfreq{n-1}}} \\
\end{array}
\right). 
\end{split}
\end{equation}
The frequencies can be obtained by the recursion formula
\begin{equation}\label{eq:recur}
\onfreq{n+1}^2 = \onfreq{n}^2\left[1 - \frac{1}{\onfreq{n}}\lie_\n\left(\frac{\donmfreq{n}}{\onfreq{n}}\right) - \frac{\donmfreq{n}^2}{\onfreq{n}^2}\right].
\end{equation}
This equation is a generalization to the well know recursion for the WKB approximation [see Eq.~(25) and (26) in~\cite{Chung2003}, for example].

The leading order adiabatic approximation in the $c_n$ representation is
\begin{align}\label{eq:lead:1}
\BR_1^{c_n} &= \frac{e^{-\ci\int\limits_{t_0}^t\onfreq{n}{}(t_1)\dd{}t_1}}{\sqrt{2\onm{n}\onfreq{n}}} + \OO{\frac{\donmfreq{n}}{\onfreq{n}}}, \\ \label{eq:lead:2}
\BR_2^{c_n} &= -\ci\sqrt{\frac{\onm{n}\onfreq{n}}{2}}e^{-\ci\int\limits_{t_0}^t\onfreq{n}(t_1)\dd{}t_1} + \OO{\frac{\donmfreq{n}}{\onfreq{n}}}.
\end{align}
Transforming them back to the $c_1$ representation, we obtain
\begin{align}
\BR_1^{c_1} &= \frac{e^{-\ci\int\limits_{t_0}^t\onfreq{n}{}(t_1)\dd{}t_1}}{\sqrt{2\onm{1}\onfreq{n}}} + \OO{\frac{\donmfreq{n}}{\onfreq{n}}}, \\ \nonumber
\BR_2^{c_1} &= -\sqrt{\frac{\onm{1}\onfreq{n}}{2}}\left(\ci+\frac{\lie_\n{\sqrt{\onm{1}\onfreq{n-1}}}}{\onfreq{n}\sqrt{\onm{1}\onfreq{n-1}}}\right)e^{-\ci\int\limits_{t_0}^t\onfreq{n}(t_1)\dd{}t_1} \\
&+ \OO{\frac{\donmfreq{n}}{\onfreq{n}}}.
\end{align}
Using the above solutions, it is easy to show that the representations $c_n$ are also unitarily equivalent to the $c_1$ representation. 

Considering the fact that all $c_n$ representations are unitarily equivalent, one can use any of them to describe a particular quantum field. The main difference between the $c_1$ and $c_n$ representations, with $n>1$, is that the latter involves a LCT whose parameters depend on both time and the eigenvalue $\EV{\HFi}$, while the transformation between the original representation and $c_1$ is only time-dependent. To interpret these new LCT physically, it is worth writing the field operator explicitly in both representations, i.e., from Eqs.~\eqref{eq:def:hatchi} and \eqref{eq:def:BF:split}, we get
\begin{equation}
\mfo^{c_n}(x) = \int\dd^3{}\HFi\HF{\HFi}\left(\BR_1^{c_n}\AO_{\HFi} + \BR_1^{c_n*}\AO_{\HFi}^\dagger\right).
\end{equation}
We omitted the basis used to define the creation and annihilation operators, because they are invariant under LCT in the Schrodinger representation. Therefore, any canonical variables $c_n$ result in the same operators, e.g., $\AO_{\HFi}[\BR^{c_1}] = \AO_{\HFi}[\BR^{c_2}]$. Expressing the field operator using $c_n$ variables and in terms of the $c_1$ representation, we obtain
\begin{equation}
\mfo^{c_n}(x) = \int\dd^3{}\HFi\HF{\HFi}\sqrt{\frac{\onm{1}\onfreq{n-1}}{\EV{\HFi}}}\fsmfo^{c_1}_{\HFi}.
\end{equation}
Introducing the following window function 
\begin{equation}\label{eq:window}
W_{c_n,c_1}(x,y) = \int\dd^3{}\HFi\HF{\HFi}(x)\HF{\HFi}(y)\sqrt{\frac{\onm{1}\onfreq{n-1}}{\EV{\HFi}}},
\end{equation}
we write the relation between the representations as
\begin{equation}
\mfo^{c_n}(x) = \int\dd^3{}y W_{c_n,c_1}(x,y)\mfo^{c_1}(y).
\end{equation}
This transformation acts as a averaging (smearing) procedure, providing the exact window function which connects to a representation of the field where the particle creation leading term is of order $2n+1$. 

It should be noted that these LCT are restricted to the modes which satisfy $\onfreq{n}^2 > 0$. In addition, the series obtained by applying Eq.~\eqref{eq:recur} is know to provide an asymptotic series approximation and, as such, it has a maximum $n$ above which the series stops converging. Thus, the window function in Eq.~\eqref{eq:window} should reflect the LCT only for the modes where the transformation is possible.\footnote{For example, one can use the identity transformation for all modes satisfying $\onfreq{n}^2 < 0$.}

In the first step to determine the $c_1$ representation, we evaluate Eq.~\eqref{eq:c1:c2} to choose $\onm{1}$ such that $\onmfreq{1}$ satisfies the necessary conditions for unitary evolution. In all subsequent steps we choose LCT where the coefficients are determined algebraically in terms of the old representation. We can instead impose that $\onm{\infty}\onfreq{\infty} = 1$, where we label this representation as $c_\infty$. This leads to a differential equation for the new frequency, i.e.,
\begin{equation}\label{eq:c1:infty}
\onfreq{\infty}^2 = \ofreq^2 - \frac{\ddot{\sqrt{\om}}}{\sqrt{\om}} + \frac{\ddot{\sqrt{\onfreq{\infty}^{-1}}}}{\sqrt{\onfreq{\infty}^{-1}}}.
\end{equation}
This equation can be easily recognized as the differential equation for the frequency $W$ in the WKB approach (see~\cite[Eq.~(3.35)]{Parker2009} for example). The main difference is that, in this case, the function $\onfreq{\infty}$ enters as the coefficient of a LCT which takes the field to a new representation where the basis functions have a simple analytic solution [since $\donmfreq{\infty} = 0$, Eqs.~\eqref{eq:AA:eom} have trivial quadrature solutions]. The ambiguity is now contained in $\onfreq{\infty}$ because, as it is defined by a differential equation, it requires the imposition of initial conditions. For example, one can choose $\onfreq{n}(t_0)$ and $\dot{\ofreq}_{c_n}(t_0)$ as initial conditions for $\onfreq{\infty}$ but, then, the same ambiguities in determining the adiabatic vacuum will persist.


\subsection{Hamiltonian Diagonalization}

The Hamiltonian for every $c_n$ representation is written in the form 
\begin{equation}
\gH^{ab} \doteq \left( \begin{array}{cc}
\onm{n}\onfreq{n}^2 & 0 \\
0 & \frac{1}{\onm{n}} \end{array} \right).
\end{equation}
The Hamiltonian operator, transformed by $\HF{\HFi}$, can be cast as [see Eq.~\eqref{eq:trans:op}]
\begin{equation}
\hat{\gH} = \frac{1}{2}\fsmvo_{\HFi,a} \gH^{ab}\fsmvo_{\HFi,b} = \frac{1}{2}\fsmvo_{\HFi,a} \SM^{ab}\gH_{a}{}^{c}\fsmvo_{\HFi,c},
\end{equation}
where in the second equality we introduced the operator $\gH_{a}{}^{c} \equiv \SM_{ab}\gH^{bc}$. The operator $\gH_{a}{}^{c}$ is Hermitian with respect to the product $\BRi_a^*\SM^{ab}\BSi_a$, i.e., $$\BRi_a^*\SM^{ab}(\gH_b{}^c\BSi_c) = (\gH_a{}^c\BRi_c)^*\SM^{ab}\BSi_b.$$ So naturally, we can diagonalize it obtaining the eigenvalues $\pm \onfreq{n}$ and eigenvectors proportional to 
\begin{equation}\label{eq:eigenvec}
\SEV_a \doteq (1, \pm \ci\onm{n}\onfreq{n}).
\end{equation}
Considering $\onfreq{n}^2 > 0$, its respective normalized vector is
\begin{equation}
\BE_a \doteq \left(\frac{1}{\sqrt{2\onm{n}\onfreq{n}}},-\ci\sqrt{\frac{\onm{n}\onfreq{n}}{2}}\right).
\end{equation}
Note that these eigenvectors are equivalent to the leading approximations given in Eqs.~\eqref{eq:lead:1} and \eqref{eq:lead:2} except for an irrelevant phase. This fact provides a connection between the instantaneous diagonalization vacuum definition (see~\cite{Mukhanov2007}, for example) and the adiabatic vacuum. The instantaneous diagonalization is equivalent to the leading order approximation in each $c_n$ canonical variables. Each one of these leading approximations $c_n$, when transformed back to the $c_1$ representation, results in the $2n$ order adiabatic vacuum. In other words, the $2n$ order adiabatic vacuum is equivalent to the diagonalization of the Hamiltonian in the $c_n$ canonical variables.

\section{Conclusions}
\label{sec:conclusions}

In this work, we studied under which conditions the evolution of a free quantum field can be represented by unitary operators. We obtained that these conditions restrict the form of the system's Hamiltonian by imposing the three requirements given by Eqs.~\eqref{eq:cond1}--\eqref{eq:cond3}. However, under a time-dependent LCT both the canonical variables and the form of the Hamiltonian are modified. This means that different choices of canonical variables lead to different Hamiltonians. Then, if the Hamiltonian does not fulfill those conditions for a given choice of canonical variables, it is possible that for other choice it may do so. Conversely, if the Hamiltonian satisfies such conditions, the new one obtained after a LCT may not do so. In short, the requirement of unitary evolution translates into the choice of canonical variables representing the system.

We evaluated this problem in Sec.~\ref{sec:back:cond}, where we assumed a general form for the mass frequency product $(\om\ofreq)^2$ of an arbitrary Hamiltonian. This form consists in a Laurent series expansion in the reciprocal Laplacian eigenvalue $\EV{\HFi}^{-1}$, whose first term is $\EV{\HFi}^{2}$. If the Hamiltonian contains only integer powers of the Laplacian, then all elements of the series expansion with odd $i$ [Eq.~\eqref{eq:def:mv}] will be zero. In this case, it is always possible to find a time-dependent LCT for which the new canonical variables allow a unitary evolution. Otherwise, if $f_{-1}(t) \neq 0$ then it is possible to find such LCT if $f_{-1}(t) = b f_{-2}(t)$ for any real constant $b$. Therefore, for most physically motivated problems, there is a LCT leading to a unique pair of canonical variables for which the new Hamiltonian satisfies the conditions [Eqs.~\eqref{eq:cond1}--\eqref{eq:cond3}], even if the original Hamiltonian does not satisfy them. 

The connection between unitary evolution, Hamiltonian and canonical variables establishes a link between the equivalence principle and unitary evolution. As we discussed in the end of Sec.~\ref{sec:back:cond}, the equivalence principle can be applied to a given quantum field representation by examining the behavior of its defining basis in the UV limit. If the basis reduces to that of quantum fields in Minkowsky space-time, then we say that this representation respects the equivalence principle. The behavior of the basis depends on both the choice of canonical variables and their initial conditions. The unique canonical pair that allows the time evolution to be implemented unitarily is the same pair that allows initial conditions such that the equivalence principle is respected. Every choice of stable representation, at least at leading order, will also respect the equivalence principle. Nevertheless, not all initial conditions satisfying the equivalence principle will be stable at some order. This shows that the equivalence principle has the same role as the requirement of unitary evolution. It can be used to select the canonical variables, and only impose asymptotic restrictions on the initial conditions of the representation basis.

Expanding the LCT group to allow time and position dependency on the LCT parameters, we deduced in Sec.~\ref{sec:higher} a recurrent series of LCT which is close related to the adiabatic vacuum conditions. This provides a physical picture for the adiabatic vacuum. The $c_n$ canonical variables are obtained by smearing the field with window functions, which depend only on the background variables. In these new canonical variables, the field evolves slowly and slowly in the sense that the time derivative of the adiabatic quantities are proportional to higher powers of $\EV{\HFi}^{-1}$. Apart from this interesting physical picture, we find out these canonical variables also useful for numerical calculation of the adiabatic basis. Usually the most computationally expensive part of the basis function analysis steams from the high frequency oscillations. In the new $c_n$ canonical variables, one can numerically solve the equations of motion using the AA variables, obtaining a fast integration time since the time derivatives of the adiabatic quantities can be made very small with an exact LCT.

Finally, these $c_n$ canonical variables also give the necessary tools to connect the vacuum choice by instantaneous Hamiltonian diagonalization with the adiabatic vacuum. Since in each $c_n$ canonical representation the Hamiltonian is modified, its eigenvectors will also be different for each choice $c_n$. We can use the eigenvector of the Hamiltonian $c_n$ to define the initial conditions in $c_1$ by simply applying the inverse LCT $c_n \rightarrow c_1$. The result is that the adiabatic vacuum of a given order in the representation $c_1$ is equivalent to the instantaneous Hamiltonian diagonalization of the $c_n$ canonical variables transformed back to the initial $c_1$ canonical pair.

\section*{ACKNOWLEDGMENTS}

S.D.P.Vitenti acknowledges financial support from Capes, under the program ``Ci\^{e}ncias sem Fronteiras'' (grant number 2649-13-6). The author also thanks Patrick Peter and Mariana Penna-Lima for their proofreading of the manuscript and the conversations regarding the topics discussed here.

\appendix

\section{Definitions and notation}
\label{app:def}

In this work we consider a background manifold containing a metric $\g_{\mu\nu}$ with signature $(-1,1,1,1)$ and inverse $\g^{\mu\nu}$. We denote the torsion-free covariant derivative compatible with this metric by $\nabla_\mu$, such that $\nabla_\mu g_{\alpha\beta} = 0$. The four dimensional natural integration form is given by 
\begin{equation}
\fint \equiv \fintc_{\mu\nu\alpha\beta}\dd{}x^\mu\wedge\dd{}x^\nu\wedge\dd{}x^\alpha\wedge\dd{}x^\beta,
\end{equation}
where $\nabla_\gamma\fintc_{\mu\nu\alpha\beta} = 0$. Given a globally defined time-like normal vector field $\n^\mu$ we can foliate the manifold and, for each hypersurface, the metric is $\h_{\mu\nu} \equiv \g_{\mu\nu} + \n_\mu\n_\nu$. The spatial covariant derivative induced in this foliation is $\scd_\mu T = \h_\mu{}^\nu\cd_\nu T$.\footnote{For general tensors, the spatial covariant derivative is obtained by projecting every index after the application of the covariant derivative.}  The constant spatial integration form for these hypersurfaces is 
\begin{equation}
\ddx = \tintz \equiv a^{-3}\tint \equiv a^{-3}\n^\mu\fintc_{\mu\nu\alpha\beta}\dd{}x^\nu\wedge\dd{}x^\alpha\wedge\dd{}x^\beta.
\end{equation}
In the above expression we used a geodesic foliation to obtain the time slices, i.e., $\n^\mu\cd_\mu\n^\nu = 0$ and the ``dot'' operator as the Lie derivative in the time direction, $\dot{T} \equiv \lie_\n T$. Given this folitation, we have a natural definition of time $t$ given by $\dot{t} = 1$ and $\scd_\mu t = 0$. The volume form $\fint$ is not constant with respect to this time direction, it is easy to show that $\lie_\n(\fint) = \cd_\mu\n^\mu\fint$. Using the fact that $\cd_\mu\n_\nu = \EC_{\mu\nu}$ is the extrinsic curvature and $\EX \equiv \EC_\mu{}^\mu$ is the expansion factor, we introduce the scale factor $a$ satisfying $\dot{a}/a = \EX/3$. Hence, it is clear that $\ddx\ddt \equiv a^{-3}\fint$ is constant, $\lie_\n(\ddx\ddt) = 0$. Here we used the notation $\ddx\ddt$ to explicit its dependency with respect to the foliation given by $\n^\mu$.

The Laplace-Beltrami operator is given by $\scd^2 \equiv \h^{\mu\nu}\scd_\mu\scd_\nu$. It scales as $a^{-2}$ for a Friedmann geometry and, therefore, we introduce its conformal version $\tilde{\scd}^2 \equiv a^2\scd^2$. The eigenfunctions of this operator are defined as,
\begin{equation}\label{eq:laplace}
\tilde{\scd}^2\HF{\HFi} = -\EV{\HFi}^2\HF{\HFi},\quad \int\limits_\ST\dd^3x \HF{\HFi_1}\HF{\HFi_2} = \dirac{3}{\HFi_1-\HFi_2},
\end{equation}
with eigenvalues $-\EV{\HFi}^2$. For flat hypersurfaces, for example, we can choose $\HF{\HFi}$ as plane waves and in this case $\HFi$ is the mode vector and $\EV{\HFi}^2 = \HFi\cdot\HFi$. Note that, since $\tilde{\scd}^2$ is constant, the eigenvalues $\EV{\HFi}^2$ are also constant.

\section{Integral Approximation}
\label{app:intapprox}

In order to calculate the approximation of the solutions, we need the series approximation of $\delta\epsilon$ and $\delta\gamma$. These functions satisfy the following integral equations,
\begin{align}
\delta\epsilon &=  2\int\limits_{t_0}^t\dd{}t_1\cos(2\bar{\varphi}_1)\dot{\omfreq}_1, \\
\delta\gamma &= 2\int\limits_{t_0}^t\dd{}t_1\frac{\sin(2\bar{\varphi}_1)}{\cosh(\epsilon_1)}\dot{\omfreq}_1,
\end{align}
while the angle $\bar{\varphi}$ satisfies [Eq.~\eqref{eq:dot:barvarphi2}]
\begin{equation}\label{eq:def:dotbarvarphi}
\dot{\bar{\varphi}} = \ofreq - \tanh(\epsilon)\sin(2\bar{\varphi}) \dot{\omfreq}.
\end{equation}
Multiplying and dividing the integrand by the frequency $\ofreq$ we obtain for $\delta\epsilon$,
\begin{equation}
\begin{split}
\delta\epsilon &= \int\limits_{t_0}^t\dd{}t_1\left\{\lie_\n[\sin(2\bar{\varphi}_1)]\frac{\dot{\omfreq}_1}{\ofreq_1} \right. \\
&\left.+ \lie_\n\left[\tanh(\epsilon_1)\frac{\sin(2\bar{\varphi})^2}{2}\right] \frac{\dot{\omfreq}_1^2}{\ofreq_1^2} + \OO{\frac{\dot{\omfreq}^3}{\ofreq^3}}\right\},
\end{split}
\end{equation}
where we used Eq.~\eqref{eq:def:dotbarvarphi} to rewrite $\ofreq$ in the numerator. The oscillatory term resulting from this substitution can be written as $\sin(4\bar{\varphi})/2$ and, therefore, it is still an oscillatory term. Due to this oscillatory nature, we can repeat the process of multiplying and dividing by $\ofreq$ to obtain higher order terms. It is easy to see that, for $\delta\epsilon$, all terms computed in this way are oscillatory. This method of approximating integrals integrating by parts is know to result in asymptotic series approximation, for more details see~\cite{Bender1978}.

Performing a similar calculation for $\delta\gamma$, we obtain
\begin{equation}
\begin{split}
\delta\gamma &= \int\limits_{t_0}^t\dd{}t_1\left\{-\lie_\n\left[\frac{\cos(2\bar{\varphi}_1)}{\cosh(\epsilon_1)}\right]\frac{\dot{\omfreq}_1}{\ofreq_1} \right. \\
&\left.- \lie_\n\left[\frac{\tanh(\epsilon_1)\cos(4\bar{\varphi})}{2\cosh(\epsilon_1)}\right]\frac{\dot{\omfreq}_1^2}{\ofreq_1^2} + \OO{\frac{\dot{\omfreq}^3}{\ofreq^3}}\right\}.
\end{split}
\end{equation}
Finally, the correction in the angle $\bar{\varphi}$ is given by the integral
\begin{equation}
\delta\bar{\varphi} = -\int\limits_{t_0}^t\dd{}t_1\tanh(\epsilon_1)\sin(2\bar{\varphi}_1)\dot{\omfreq}_1.
\end{equation}
For this equation, applying the procedure described above results in
\begin{equation}
\begin{split}
\delta\bar{\varphi} &= \int\limits_{t_0}^t\dd{}t_1\left\{\lie_\n\left[\frac{\tanh(\epsilon_1)\cos(2\bar{\varphi}_1)}{2}\right]\frac{\dot{\omfreq}_1}{\ofreq_1} - \frac12\frac{\dot{\omfreq}_1^2}{\ofreq_1}\right. \\
&\left.+ \lie_\n\left[\frac{(\tanh(\epsilon_1)^2-2)\cos(4\bar{\varphi})}{4}\right]\frac{\dot{\omfreq}_1^2}{\ofreq_1^2} + \OO{\frac{\dot{\omfreq}^3}{\ofreq^3}}\right\}.
\end{split}
\end{equation}
In this case, one of the generated terms, $\dot{\omfreq}^2/\ofreq$, is non-oscillatory. This means that this is actually a first order term, since there is no other oscillatory term to average it out. This is the reason why one needs to calculate the expansion up to second order: one must check if there are non-oscillatory terms contributing to the first order part of the series.

Integrating by parts the first order term of $\delta\epsilon$, we obtain
\begin{equation}\label{eq:approx:de}
\delta\epsilon \approx \left.\sin(2\bar{\varphi}_1)\frac{\dot{\omfreq}_1}{\ofreq_1}\right\vert_{t_0}^t - \int\limits_{t_0}^t\dd{}t_1\sin(2\bar{\varphi}_1)\lie_\n\left(\frac{\dot{\omfreq}_1}{\ofreq_1}\right).
\end{equation}
Note that the second term on the right hand side is again the integral of an oscillatory function times a function of the background, as the original integral for $\delta\epsilon$. Nonetheless, in this case it is already of order $\ofreq^{-1}$. Therefore, performing the same approximation scheme, its contribution will be of order $\ofreq^{-2}$. In short, we need only the first term in the integration by parts to obtain a first order approximation for each variable, i.e.,
\begin{align}\label{eq:approx:dg}
\delta\gamma &\approx \left.-\frac{\cos(2\bar{\varphi}_1)}{\cosh(\epsilon_1)}\frac{\dot{\omfreq}_1}{\ofreq_1}\right\vert_{t_0}^t, \\ \label{eq:approx:df}
\delta\bar{\varphi} &\approx \left.\frac{\tanh(\epsilon_1)\cos(2\bar{\varphi}_1)}{2}\frac{\dot{\omfreq}_1}{\ofreq_1}\right\vert_{t_0}^t - \int\limits_{t_0}^t\dd{}t_1\frac{\dot{\omfreq}_1^2}{2\ofreq_1}.
\end{align}
All omitted terms in the above equations are of order $\ofreq^{-2}$ or higher.

\end{document}